\input jytex.tex   
\prelabel {af1}{\hbox {20}}
\prelabel {app1}{\hbox {50}}
\prelabel {app2}{\hbox {51}}
\prelabel {app3}{\hbox {52}}
\overfullrule=0cm
\typesize=10pt
\magnification=1200
\baselineskip17truept
\hsize=6truein\vsize=8.5truein
\sectionnumstyle{blank}
\chapternumstyle{blank}
\chapternum=1
\sectionnum=1
\pagenum=0

\def\begintitle{\pagenumstyle{blank}\parindent=0pt\begin{narrow}[0.4in]}
\def\endtitle{\end{narrow}\newpage\pagenumstyle{arabic}}


\def\beginexercise{\vskip 20truept\parindent=0pt\begin{narrow}[10 
truept]}
\def\endexercise{\vskip 10truept\end{narrow}}


\def\eql#1{\eqno\eqnlabel{#1}}
\def\ref{\reference}
\def\peq{\puteqn}
\def\pref{\putref}

\def\mgn{\marginnote}
\def\bex{\begin{exercise}}
\def\eex{\end{exercise}}


\font\open=msbm10 
\def\mbox#1{{\leavevmode\hbox{#1}}}

\def\hspace#1{{\phantom{\mbox#1}}}
\def\oR{\mbox{\open\char82}}

\def\al{\alpha}

\def\ga{\gamma}

\def\Ga{\Gamma}

\def\ep{\epsilon}

\def\la{\lambda}
\def\La{\Lambda}

\def\Om{\Omega}

\def\Si{\Sigma}

\def\ze{\zeta}

\def\De{\Delta}

\def\zf{$\zeta$--function}


\def\frac#1/#2{\leavevmode\kern.1em
\raise.5ex\hbox{\the\scriptfont0 #1}\kern-.1em/\kern-.15em
\lower.25ex\hbox{\the\scriptfont0 #2}}
\def\sfrac#1/#2{\leavevmode\kern.1em
\raise.5ex\hbox{\the\scriptscriptfont0 #1}\kern-.1em/\kern-.15em
\lower.25ex\hbox{\the\scriptscriptfont0 #2}}

\def\gtorder{\mathrel{\raise.3ex\hbox{$>$}\mkern-14mu
             \lower0.6ex\hbox{$\sim$}}}
\def\ltorder{\mathrel{\raise.3ex\hbox{$<$}\mkern-14mu
             \lower0.6ex\hbox{$\sim$}}}

\def\semidirprod{\rlap{\ss C}\raise1pt\hbox{$\mkern.75mu\times$}}
\def\for{\lower6pt\hbox{$\Big|$}}
\def\fish{\kern-.25em{\phantom{abcde}\over \phantom{abcde}}\kern-.25em}


\def\boxit#1{\vbox{\hrule\hbox{\vrule\kern3pt
        \vbox{\kern3pt#1\kern3pt}\kern3pt\vrule}\hrule}}
\def\dalemb#1#2{{\vbox{\hrule height .#2pt
        \hbox{\vrule width.#2pt height#1pt \kern#1pt
                \vrule width.#2pt}
        \hrule height.#2pt}}}

\def\frac#1#2{{{#1}\over{#2}}}


\def\eg{{\it e.g. }}
\def\ie{{\it i.e. }}

\def\pa{\partial}


  %

\def\3j#1#2#3#4#5#6{\left\lgroup\matrix{#1&#2&#3\cr#4&#5&#6\cr}
\right\rgroup}

\def\man{{\cal M}}

\def\can{{\cal N}}

\def\m?{\mgn{?}}


\def\pa{\partial}

\def\beq{\begin{eqnarray}}
\def\eeq{\end{eqnarray}}

\def\zn{\zeta_{{\cal N}}\left(}

\def\g{\Gamma\left(}

\def\cab{{\cal B}}
\def\cam{{\cal M}}
\def\can{{\cal N}}

\def\summenp{\sum d(\nu) \sum_{m=0}^{p-1} \sum_{l=0}^m (-1)^m \delta_{ml}^{p}}


\def\aop#1#2#3{{\it Ann. Phys.} {\bf {#1}} (19{#2}) #3}

\def\cmp#1#2#3{{\it Comm. Math. Phys.} {\bf {#1}} (19{#2}) #3}
\def\cqg#1#2#3{{\it Class. Quantum Grav.} {\bf {#1}} (19{#2}) #3}

\def\jgp#1#2#3{{\it J. Geom. and Phys.} {\bf {#1}} (19{#2}) #3}
\def\jmp#1#2#3{{\it J. Math. Phys.} {\bf {#1}} (19{#2}) #3}
\def\jpa#1#2#3{{\it J. Phys.} {\bf A{#1}} (19{#2}) #3}

\def\np#1#2#3{{\it Nucl. Phys.} {\bf B{#1}} (19{#2}) #3}
\def\pl#1#2#3{{\it Phys. Lett.} {\bf B{#1}} (19{#2}) #3}

\def\prD#1#2#3{{\it Phys. Rev.} {\bf D{#1}} (19{#2}) #3}

\def\cras#1#2#3{{\it Comptes Rend. Acad. Sci. (Paris)} {\bf{#1}} (#2) #3}

\def\mpcps#1#2#3{{\it Math. Proc. Camb. Phil. Soc.} {\bf{#1}} (19{#2}) #3}

\def\am#1#2#3{{\it Acta Mathematica} {\bf {#1}} (19{#2}) #3}
\def\aim#1#2#3{{\it Adv. in Math.} {\bf {#1}} (19{#2}) #3}
\def\ajm#1#2#3{{\it Am. J. Math.} {\bf {#1}} ({#2}) #3}

\def\aom#1#2#3{{\it Ann. of Math.} {\bf {#1}} (19{#2}) #3}

\def\cpde#1#2#3{{\it Comm. Partial Diff. Equns.} {\bf {#1}} (19{#2}) #3}

\def\invm#1#2#3{{\it Invent. Math.} {\bf {#1}} (19{#2}) #3}
\def\ijpam#1#2#3{{\it Ind. J. Pure and Appl. Math.} {\bf {#1}} (19{#2}) #3}
\def\jdg#1#2#3{{\it J. Diff. Geom.} {\bf {#1}} (19{#2}) #3}

\def\jmpa#1#2#3{{\it J. Math. Pures. Appl.} {\bf {#1}} ({#2}) #3}

\def\ojm#1#2#3{{\it Osaka J.Math.} {\bf {#1}} ({#2}) #3}

\def\pja#1#2#3{{\it Proc. Jap. Acad.} {\bf {A#1}} (19{#2}) #3}

\def\tams#1#2#3{{\it Trans. Am. Math. Soc.} {\bf {#1}} (19{#2}) #3}

\begin{title}  
\vglue 1truein
\vskip15truept
\centertext {\Bigfonts \bf Smeared heat-kernel coefficients on the}
\vskip3truept
\centertext{\Bigfonts \bf ball and generalized cone}
\vskip 20truept 
\centertext{J.S.Dowker\footnote{dowker@a3.ph.man.ac.uk},\quad
Klaus Kirsten\footnote{Klaus.Kirsten@itp.uni-leipzig.de}} 
\vskip 7truept
\centertext{*\it Department of Theoretical Physics,\\
The University of Manchester, Manchester, England}
\vskip10truept
\vskip10truept
\vskip7truept
\centertext{\dag\it Universit\"at Leipzig, Institut f\"ur Theoretische 
Physik,\\ Augustusplatz 10, 04109 Leipzig, Germany}
\vskip 20truept
\centertext {Abstract}
\vskip10truept
\begin{narrow}
We consider smeared zeta functions and heat-kernel coefficients on the 
bounded, generalized cone in arbitrary dimensions. The specific case of 
a ball is analysed in detail and used to restrict the form 
of the heat-kernel coefficients $A_n$ on smooth manifolds with boundary. 
Supplemented by conformal transformation techniques, it is used to provide 
an effective scheme for the calculation of the $A_n$. As an 
application, the complete $A_{5/2}$ coefficient is given. 
\end{narrow}
\vskip 5truept
\vskip 60truept
\vfil
\end{title}
\pagenum=0
\section{\bf 1. Introduction}
The coefficients $A_n$ in the small-time asymptotic expansion 
of the heat-kernel corresponding to a Laplacian-like operator on smooth 
manifolds (possibly with a boundary)
play important roles both in quantum field
theory and pure mathematics. 

Many schemes for their evaluation have been developed which 
may be divided roughly into ``direct" 
[\pref{ivan91,fulken88,amster89,osb91,and92,trento90,ven97}] and 
``indirect" [\pref{BandG2,BGV,Moss,MandD,DandS}]. 
For manifolds without boundary, the $A_n$ 
are determined by algebraic equations and their 
computation can be, and has been, done by computer. In principle the only
things needed are the  coincidence limits of the 
geodesic distance and of its derivatives at two points. Methods employing one 
or another variation on this scheme can be termed direct. 
The ``indirect" method has been developed 
most systematically by Branson and Gilkey [\pref{BandG2}]. Conformal 
transformation techniques give relations between the numerical multipliers
in the heat-kernel coefficients. However, on its 
own, this method is unable to determine the coefficients fully.
Additional information is needed, coming from other functorial relations
or special case calculations [\pref{BandG2}]. Given a subset of numerical 
coefficients the method then provides  all required information with 
relative ease. The aim of the present article is to give 
a special case calculation containing enough information which, when
supplemented by the methods of [\pref{BandG2}], leads to a very 
effective scheme for the evaluation of at least a substantial part of any 
$A_n$. 

In a previous article [\pref{BKD}] (see also [\pref{BEK}]) 
we considered the non-smeared heat-kernel 
of the Laplacian with Dirichlet or Robin boundary conditions on the 
$(d+1)$-dimensional bounded cone. Here we  
generalize this work to include a smearing function. This is an essential step
in elucidating the form of the coefficients in the presence of boundaries 
and is also vital when conformal properties are being analyzed, particularly
those of the functional determinant. 

The resulting restrictions are more informative than others available and are 
used to determine the first few heat-kernel coefficients on an 
arbitrary smooth manifold with boundary. The complete pure boundary 
coefficient $A_{5/2}$ is given showing the practicability 
of our approach for high orders [\pref{KK}]. 

The paper is organized as follows. 
In section 2 a method is developed for the calculation of the smeared 
heat-kernel
coefficients on the generalized cone. Both Dirichlet and Robin boundary 
conditions are treated. This information is used in section 3 to put 
restrictions on the general form of the coefficients. 
Supplemented by additional relations [\pref{BandG2}], the 
coefficients $A_0,...,A_{5/2}$, are (re)considered and fully determined. The 
Conclusion summarizes our main results and suggests extensions.

\section{\bf 2. Smeared  \zf\ on the generalized cone}
Our immediate objective is the determination  of the smeared heat-kernel 
coefficients on the $(d+1)$-dimensional bounded generalized cone $\cam 
=I\times\can$ with the hyperspherical metric {\it cf} [\pref{cheeg1}] 
$$
ds^2=dr^2+r^2d\Si^2,  \eql{2.1}
$$
where $d\Si^2$ is the metric on the manifold $\can$, and $r$ runs from $0$
to $1$.
$\can$ will be referred to as the base of the cone. 
If it has no boundary then it is the boundary of $\man$ with 
extrinsic curvature $K^a_b = \delta ^a _b $. 

We consider the Laplacian on $\cam$, 
$$
\De_{\cam}={\pa^2\over\pa r^2}+{d\over r}{\pa\over\pa r}
+{1\over r^2}\De_{\can},  \eql{2.2}
$$
with Dirichlet or Robin boundary conditions. The nonzero 
eigenmodes of $\De_\cam$ that are finite at the origin have
eigenvalues $-\al^2$ and are of the form
$$
{J_{\nu}(\al r)\over r^{(d-1)/2}}\,Y(\Om), 
\eql{2.3}
$$
where the harmonics on $\can$ satisfy
$$
\De_{\can} Y(\Om)=-\la^2 Y(\Om)
\eql{2.4}
$$
and 
$$
\nu^2= \la^2+(d-1)^2/4  .
\eql{2.5}
$$
One can also add the coupling $-\xi R$ to
$\De_\man$, changing the following analysis slightly [\pref{BKD}].

In order to deal with the smeared \zf\ we parallel the 
analysis presented in [\pref{BKD}] and refer to it for further details.
We consider both Dirichlet, 
$$
J_{\nu}(\al) =0, 
\eql{2.6}
$$
and Robin boundary conditions,
$$
\left(1-\frac D 2 -S \right) J_{\nu} (\al ) +\al J' _{\nu} (\al ) =0,
\eql{2.7}
$$
where $D=d+1$. Pure Neumann conditions correspond to $S =0$. 

Our main interest is in the calculation of the boundary terms of the 
heat-kernel expansion.
A convenient way of handling these is to introduce 
smeared, integrated quantities \eg the heat-kernel
$$
K(F;\tau)=\int dx\, F(x) K(x,x,\tau)
\eql{2.8}
$$
and its Mellin transform, the \zf,
$$
\ze(F;s)=\sum_\al\int dx\, F(x)\phi(x)\phi^*(x)\,{1\over \al^{2s}}
\eql{smzet}
$$
in terms of eigenfunctions, $\phi$, and eigenvalues, $-\al^2$.
In addition, the base \zf\ 
$$
\zeta_{\can }(s) = \sum_\nu d(\nu ) \nu ^{-2s} \eql{2.10}
$$
will turn out to be very useful. Here, $d(\nu)$ is the `angular' degeneracy.

On the generalised cone, the eigenfunctions (\peq{2.3}) are products of 
Bessel functions and\mgn{$\pa\can=\emptyset$}
`spherical', \ie base, harmonics. If we smear in the radial coordinate only, 
then in (\peq{smzet}) 
the integration over the base yields exactly the same degeneracies
as in the unsmeared case, \ie the $d(\nu )$, and the contour expression
for the \zf\ on $\man$ reads (we treat Dirichlet scalars first)
$$
\ze(F;s)=\sum\int_\ga {dk\over2\pi i}k^{-2s}\int_0^1 dr\,F(r)\bar
J^2_\nu(kr) r\,d(\nu){\pa\over\pa k}\ln J_\nu(k)
\eql{smsc}$$
where the bar stands for normalised,
$\bar J_\nu (\alpha r) = \sqrt 2 J_\nu (\al r ) / J' _\nu (\al )$. 

The boundary parts of the coefficients contain normal derivatives,
$F_{r \ldots r}$, of $F$ and we choose for $F$ a polynomial in $r^2$ 
(why $r^2$ will be clear soon) that
contains sufficient independent derivatives to pick out the relevant 
contributions. For example, in 
$A_1$, since there is only one normal derivative $F_r$, it is sufficient 
to take
$$
F(r)=f_0+f_1\,r^2
\eql{f1}
$$
and to use 
$$
F(1) = f_0 +f_1 \qquad F_r( 1) = 2 f_1
\eql{f2}
$$
in order to identify the boundary terms.

To explain our method more precisely, we continue with this simple example, 
(\peq{f1}), and afterwards generalize to an arbitrary polynomial. Using
$$
\int_0^1 dr \,\, r^3 \bar J _\nu ^2 (\al r ) 
= \frac 2 3 \frac{\nu ^2 -1} {\al  ^2} + \frac 1 3 
\eql{eff2}
$$
and substituting (\peq{f1}) into (\peq{smsc})
we obtain two contributions,
$$
\eqalign{
\zeta_{\cam} (F;s) =& (f_0 +\frac 1 3 f_1 ) \zeta_{\cam } (s) \cr
 & +\frac 2 3 f_1 \sum (\nu ^2 -1) d(\nu ) 
\int_\ga {dk\over2\pi i}k^{-2(s+1)}{\pa\over\pa k}\ln J_\nu(k)\cr }.
\eql{bp}
$$
Here, $\zeta_{\cam } (s)$ is defined to be $\zeta_{\cam} (1;s)$
and is known from our previous analysis [\pref{BKD}].
Also the second line in (\peq{bp}) may be 
immediately given 
by direct comparison with our previous calculation.
The contour integral is the same as previously apart from 
replacing $s\to s+1$.
For the second term this is already all we need. The first term contains
a factor $\nu^2$, raising the argument of the base zeta function by one.
(See equation (\peq{af1})).

In order to describe the method further, it is necessary to 
use some notation introduced in [\pref{BKD}]. 
For the calculation of the heat-kernel coefficients, a split of the 
zeta-function into two parts is very useful. One part contains all the 
relevant contributions and comes from the uniform asymptotic expansion 
of the Bessel function $I_\nu (k)$. As has been shown in [\pref{BKD}] 
these are the only contributions to the heat-kernel coefficients. 
Explicitly, for $\nu \to \infty$ with $z=k/\nu$
fixed one has,

$$
I_{\nu} (\nu z) \sim \frac 1 {\sqrt{2\pi \nu}}\frac{e^{\nu
\eta}}{(1+z^2)^{\frac 1 4}}\left[1+\sum_{k=1}^{\infty} \frac{u_k (t)}
{\nu ^k}\right],
\eql{3.3}
$$
with $t=1/\sqrt{1+z^2}$ and $\eta =\sqrt{1+z^2}+\ln
\big(z/(1+\sqrt{1+z^2})\big)$.
Any required number of $u_k (t)$ polynomials can be obtained via the 
recursion relation given in [\pref{AandS,olver54}]. 
In addition, we need the cumulant
expansion
$$
\ln \left[1+\sum_{k=1}^{\infty} \frac{u_k (t)}{\nu ^k}\right] \sim
\sum_{n=1}^{\infty} \frac{D_n (t)}{\nu ^n}
\eql{3.4}
$$
where the $D_n$ have the polynomial structure
$$
D_n (t) = \sum_{b=0}^n x_{n,b}\, t^{n+2b}.
\eql{3.5}
$$
The second part of the split, named $Z(s)$ in [\pref{BKD}] and 
accordingly $Z(F;s)$ here, is analytic and is of no relevance to 
the construction of the coefficients. 

By adding and subtracting $L$ leading terms of the asymptotic
expansion, (\peq{3.4}), and performing the same steps as described
in [\pref{BKD}] one finds the aforementioned split
$$
\zeta_{\cam} (F;s) =Z(F;s) +\sum_{i=-1}^{L}A_i(F;s),
\eql{3.6}
$$
with the definitions
$$
\eqalign{
A_{-1} (F;s) &= \frac 1 {4\sqrt{\pi}} \frac{\Gamma \left(s-\frac 1 2\right)}
{\Gamma (s+1)} \zn  s-1/2 \right) \left[f_0 +\frac 1 3 f_1  +\frac 2 3 
f_1 \frac {s-1/2} {s+1} \right] \cr
& -\frac 2 3 f_1 \frac 1 {4\sqrt{\pi}}
\frac{\Gamma \left(s+\frac 1 2\right)}
{\Gamma (s+2)} \zn  s+1/2 \right)\cr
A_0 (F;s) &=  -\frac 1 4 \zn s\right) \left[f_0+f_1\right ] -\frac 1 4
 \zn s+1\right) f_1 \cr
A_i (F;s) &= -\frac 1 {\Gamma (s)} \zn  s+i/2 \right)
\sum_{b=0}^i x_{i,b} \frac{\g s+b+i/2\right)}
{\g b+i/2\right)}\cr
& \hspace{ -\frac 1 {\Gamma (s)} \zn  s+i/2 \right) \sum_{b=0}^i  }
\left[ f_0 +\frac 1 3 f_1 +\frac 2 3 f_1 \frac{s+b+i/2} s \right] \cr
& -\frac 2 3 f_1 \zn s+1+i/2 \right)
\sum_{b=0}^i x_{i,b} \frac{\g s+1+b+i/2\right)}
{\Gamma (s+1)\g b+i/2\right)}\cr}
\eql{af1}
$$
As is apparent in (\peq{af1}), base contributions
are separated from radial ones. This enables 
the heat-kernel coefficients of the Laplacian on the manifold $\cam$ to be
written in terms of those on $\can$.

In the next section we discuss the restrictions our calculation
places on the general form of the heat-kernel coefficients. 
It is known, for example, that the coefficient $A_2$ contains the third 
normal derivative of the smearing function $F$ and the higher coefficients 
involve  correspondingly higher
derivatives. It is thus obvious that the $F(r)$ employed earlier (\peq{f1})  
will not be
general enough to discuss coefficients beyond $A_{3/2}$. In order to apply
our technique to all higher coefficients, at least in principle, we 
consider the polynomial 
$$
F(r)=\sum_{n=0}^N f_n r^{2n}.
\eql{pol1}
$$
This leads to normalization integrals of the type
$$\eqalign{
S[1+2p]=
\int_0^1dr\,\bar J_\nu^2(\al r)r^{1+2p},
\cr}
\eql{defschaf}$$
which can be treated using Schafheitlin's reduction formula
[\pref{watson}]. Writing this formula for the case
when $\al$ is a zero of the Bessel function, $J_\nu(\al)=0$, one has
$$
\int_0^1dr\,\bar J_\nu^2(\al r)r^{\mu+2}={\mu+1\over \mu+2}{\big(\nu^2-
(\mu+1)^2/4\big)\over \al^2}\int_0^1dr\,\bar J_\nu^2(\al r)r^\mu
+{1\over\mu+2}.
\eql{reduc}$$
This can be iterated down to the standard normalisation value, $\mu=1$, and is
the origin of (\peq{eff2}). In 
order to use this formula, which is our essential technical novelty, 
we see that it is necessary to have a polynomial in $r^2$.

Schafheitlin's formula gives the recursion for the 
normalization integrals (\peq{defschaf}),
$$\eqalign{
S[1+2p]={2p \over 2p+1} {\nu^2 -p^2 \over \alpha^2} S[2p-1] +{1 \over
2p+1} \cr}
\eql{schaf}$$
so that $S[1+2p]$ has the following form,
$$\eqalign{
S[1+2p]=\sum_{m=0}^p \left( {\nu \over \alpha} \right) ^{2m}
\sum_{l=0}^m \gamma_{ml}^p \nu^{-2l} \cr}
\eql{machine}$$
with the numerical coefficients $\gamma_{ml}^p$ being easily determined 
recursively.

As seen in the treatment of the function $F(r)$ in equation 
(\peq{f1}), using the same rules of replacement, the $A_i(s)$ read, after
some rearrangement,
\begin{ignore}
$$\eqalign{
A_{-1} (F;s)  = &{1\over 4\sqrt{\pi}  }
\sum_{p=0}^N f_p \sum_{m=0}^p \sum_{l=0}^m \gamma_{ml}^p
 {\Gamma (s-1/2 +m)  \over
\Gamma (s+1+m)} \zn s- {1\over 2}+l \right) \cr
A_0 (F;s) =& -{1\over 4}
\sum_{p=0}^N f_p \sum_{m=0}^p \sum_{l=0}^m \gamma_{ml}^p
\zn s+l\right) \cr
A_i (F;s) =& -
\sum_{p=0}^N f_p \sum_{m=0}^p \sum_{l=0}^m \gamma_{ml}^p
\zn s+{i \over 2} +l\right) \cr
& \sum_{b=0}^i x_{i,b} {\Gamma (s+b+i/2+m) \over \Gamma (s+m) \Gamma (b+i/2)}
\cr}
\eql{as2}$$

By interchanging the summations (\peq{as2}) can be rewritten
\end{ignore}
$$\eqalign{
A_{-1}(F;s)=&{1\over4\sqrt\pi}\sum_{l=0}^N\bigg[\sum_{m=l}^N L^{(N)}_{m,l}
{\Ga\big(s-1/2+m\big)\over\Ga\big(s+1+m\big)}\bigg]\ze_\can\big(s-1/2+l\big)
\cr
A_0(F;s)=&-{1\over4}\sum_{l=0}^N\bigg[\sum_{m=l}^NL^{(N)}_{m,l}\bigg]
\ze_\can(s+l)\cr
A_i(F;s)=&-\sum_{l=0}^N\bigg[\sum_{m=l}^NL^{(N)}_{m,l}
 \sum_{b=0}^i x_{i,b} {\Gamma (s+b+i/2+m) \over \Gamma (s+m) \Gamma (b+i/2)}
\bigg]\ze_\can\big(s+l+i/2\big)\cr}
\eql{as3}$$
where the linear form in the $f_p$ is defined by
$$
L^{(N)}_{m,l}=\sum_{p=m}^N \ga^p_{ml}\,f_p.
$$

For Dirichlet boundary conditions,
these formulas provide the generalization of our formalism [\pref{BKD}] 
to the radially smeared case. This is enough for our purposes
because the general forms of the heat-kernel coefficients contain only 
normal derivatives and these are radial derivatives on the generalized cone.

In the special case of the $D$--ball, 
the residues of the poles of the base (\ie sphere) \zf\ are given in terms 
of Bernoulli polynomials and the ball coefficients are then efficiently 
evaluated 
by machine [\pref{BKD}]. One could equally well take the torus as the base 
manifold, but the information obtained differs only slightly.

We now turn to Robin boundary conditions.
It is possible to proceed in the same way as for Dirichlet but complications
arise and the situation is sufficiently different so as to warrant a separate
treatment.

Write the Robin condition ({\peq{2.7}), as
$$
G_\nu(\al)=\al J'_\nu(\al)+uJ_\nu(\al)=0
$$

The normalisation is
$$
\int_0^1 J_\nu^2(\al r)rdr={1\over2\al^2}\big(\al^2-\nu^2+u^2\big)J_\nu^2(\al)
\eql{rnorm}$$
and the normalised Schafheitlin formula reads
$$\eqalign{
\int_0^1dr\,\bar J_\nu^2(\al r)r^{\mu+2}&={\mu+1\over \mu+2}{\big(\nu^2-
(\mu+1)^2/4\big)\over \al^2}\int_0^1dr\,\bar J_\nu^2(\al r)r^\mu \cr
& +{1\over\mu+2}\bigg(1+{(\mu+1)(u+{1\over2}(\mu+1))\over\al^2-\nu^2+u^2}
\bigg).
\cr}
\eql{reduc2}
$$
Continuing as in the Dirichlet case, and defining $S[1+2p]$ as in
(\peq{defschaf}), we find the reduction formula
$$\eqalign{
S[1+2p]&={2p \over 2p+1} {\nu^2 -p^2  \over \alpha ^2} S[2p-1] \cr
   &+ {1 \over 2p+1} \left( 1+{2p (u+p) \over \alpha^2 +u^2 -\nu ^2 }
\right).   \cr} \eql{defschafrob}
$$
Explicitly this gives the following form
$$\eqalign{
S[1+2p] &= \sum_{m=0}^p \left({\nu \over \alpha} \right) ^{2m}
\sum_{l=0}^m \gamma _{ml}^p\, \nu ^{-2l} \cr
& +{1 \over \alpha^2 +u^2 -\nu ^2} \sum_{m=0}^{p-1} \left(
{\nu \over \alpha }\right) ^{2m} \sum_{l=0}^n \delta _{ml} ^p\, \nu^{-2l} \cr}
\eql{machinerob}
$$
where the $\gamma_{ml}^p$ are the same as in (\peq{machine})
and the $\delta_{ml}^p$ are also easily determined by machine.

For the \zf\ we have
$$
\ze^{\rm Rob}(F;s)=\sum\int_\ga {dk\over2\pi i}k^{-2s}\int_0^1 dr\,F(r)\bar
J^2_\nu(kr) r\,d(\nu){\pa\over\pa k}\ln G_\nu(k)
\eql{smsc2}$$
where the contour $\gamma$ has to be chosen so as to enclose  the 
zeros of {\it only} $G_{\nu} (k)$. Thus the poles of $S[1+2p]$, located at
$k=\pm \sqrt{\nu^2 -u^2}$, must be outside the contour. It is
important to locate the contour properly because, when deforming it to
the imaginary axis, contributions from the pole at $k=\sqrt{\nu^2-u^2}$ arise.

As a result, apart from contributions identical to (\peq{as3}), with the
usual changes between Dirichlet and Robin boundary conditions
[\pref{BKD}], we have the extra pieces
$$\eqalign{
\zeta_{\delta} ^{p} (F;s) &= \frac {\sin \pi s }{\pi}
\sum d(\nu) \sum_{m=0}^{p-1} \sum_{l=0}^m  \delta_{ml}^{p}\, \nu^{-2l}
\cr
& \int_{m/\nu}^{\infty}
dz \,\, \frac{[(z\nu)^2 -m^2 ]^{-s} }
{u^2 -\nu^2 (1+z^2)} z^{-2m}
\frac{\partial}{\partial z} \ln \big(uI_{\nu} (z\nu) +z\nu
I_{\nu} ' (\nu z) \big)
\cr} \eql{zetarobin}
$$
$$\eqalign{
\zeta_{\rm shift}^p (F;s) &= -\frac {1}{ 2} \sum_{m=0}^{p-1}
\sum_{l=0}^m \delta_{ml}^p \sum d(\nu ) \nu^{2m-2l} (\nu ^2 -u^2 )
^{-s-m-1/2} \cr
&\quad\quad \frac{\partial}{\partial k} \ln\big(kJ_{\nu}' (k) +
uJ_{\nu} (k)\big) |_{k=\sqrt{\nu^2 -u^2} }\cr}
\eql{robshift}
$$
the last one arising on moving the contour over the pole at
$k=\sqrt{\nu^2 -u^2}$. These are the contributions
additional to those in the Dirichlet case.
The index $p$ refers to the fact that these are the contributions coming
from the power $r^p$ in (\peq{pol1}). In order to obtain the full zeta 
function, the $\sum_{p=0}^N f_p \zeta^p$ has to be done.

Looking at (\peq{zetarobin}), we first define the
asymptotic contributions $A_{i,\delta}^p (F;s)$ in the same manner
as before by taking the different terms in the asymptotic expansion
of the argument of the logarithm. We illustrate the calculation by dealing with
$$\eqalign{
A_{-1,\delta} ^{p} (F;s) &= \frac{\sin \pi s} {\pi}
\summenp \nu^{1-2l} \cr
& \int_{m/\nu} ^{\infty} dz\,\,
\frac{[(z\nu) ^2 -m^2 ] ^{-s} }
{(u^2 - \nu^2 (1+z^2) z^{2m+1} } (1+z^2)^{1/2}. \cr} 
\eql{adelta}
$$
Using the expansion for small $u$,
$$
\frac 1 {u^2-\nu^2 (1+z^2) } = -\sum_{i=0}^{\infty}
\frac{u^{2i} }{ (\nu^2 ) ^{i+1} (1+z^2 )^{i+1} }
$$
one arrives at
$$\eqalign{
A_{-1,\delta} ^{p} (F;s) &=- \frac{\sin \pi s} {\pi}
\sum_{i=0} ^{\infty} u^{2i} \summenp\, \nu^{-1-2i-2l} \cr
& \int_{m/\nu} ^{\infty} dz\,\,
\frac{[(z\nu) ^2 -m^2 ] ^{-s} }
{(1+z^2)^{i+1/2} z^{2m+1} }.  \cr}
\eql{adeltaexp}
$$
At this point we can continue as in previous articles by realizing that
the above integrals are representations of a hypergeometric function
[\pref{GandR}]. With the help of their Mellin-Barnes
representation [\pref{GandR}] the $\nu$-summation can be done, 
yielding the base
\zf, and in the massless case our final result reads
$$\eqalign{
A_{-1,\delta }^{p} (F;s) &= \frac 1 {2\Gamma (s)} \sum_{ i=0} ^{\infty}
u^{2i} \sum_{m=0}^{p-1} \sum_{l=0} ^m \delta _{ml} ^{p} \cr
& \frac{\Gamma (-s-m)
\Gamma (s+i+m+1/2)  } {\Gamma (-s+1) \Gamma (i+1/2) }
\zeta_{{\cal N}} (s+l+i+1/2). \cr}
\eql{adeltafinal}
$$

In the same way one obtains for the other $A_{i,\delta}^p (s)$,
$$\eqalign{
A_{0,\delta}^p (F;s) &= -\frac{1} {4\Gamma (s)} \sum_{i=0}^{\infty}
\frac{u^{2i}}{\Gamma (i+2)} \sum_{m=0}^{p-1} \sum_{l=0}^m (-1)^m\delta_{ml}^p
\cr
&\frac{\Gamma  (s+i+m+1) \Gamma (1-s-m)}{\Gamma (1-s)}
\zn s+i+l+1\right)
\cr}
\eql{anullfinal}
$$
$$\eqalign{
A_{n,\delta}^p (F;s) &=\frac 1 {2\Gamma (s)} \sum_{i=0}^{\infty} u^{2i}
\sum_{m=0}^{n-1} \sum_{l=0}^m (-1)^m\delta_{ml}^p\sum_{b=0}^n x_{n,b} (n+2b)\cr
&\frac{\Gamma (1-s-m) \Gamma(s+i+n/2+b+m+1)}{\Gamma(1-s)\Gamma(i+n/2+b+2)}
\zn s+i+l+1+n/2\right) \cr}
\eql{anfinal}
$$
These forms are well suited for machine evaluation  and the residues relevant
for the heat-kernel expansion are thereby quickly determined.

The remaining task is to deal with $\zeta_{\rm shift}^p (F;s)$ defined in
(\peq{robshift}). To get the relevant residues we need the asymptotic
behaviour of $J_{\nu} $, information on which can be
found in Abramowitz and Stegun [\pref{AandS}]. 
Ultimately, as a practical application, we want to restrict the general form 
of the $A_{5/2}$ coefficient and so, restricting the calculation 
to the order necessary for this coefficient, we arrive at
$$\eqalign{
\zeta_{\rm shift}^p (F; s) &= -\frac 1 2 \sum_{m=0}^{p-1} \sum_{l=0}^m
\delta_{ml} ^p\times \cr
 & \left( u\zn s+l+1\right) +u^3 (s+m+1) \zn s+l+2 \right) +...\right)
\cr}
\eql{finalshift}
$$
after some algebra.
As mentioned, the dots indicate contributions having their rightmost pole
to the left of $s=(D-5)/2$.

All the relevant results for the calculation up to the $A_{5/2}$ 
coefficient are now to hand. It would be possible to go further, if desired.
(But see our cautionary note at the end.) 

\section{\bf 3. Heat-kernel coefficients on general manifolds}
In this section we describe the restrictions placed on the
general form of the heat-kernel coefficients by our special case evaluation.
Because the case we treat has vanishing Riemann tensor
and constant extrinsic curvature, it cannot, in general, 
determine the complete coefficient. However, supplemented by 
a lemma on product manifolds and using relations of the heat-kernel 
coefficients under conformal rescalings [\pref{BandG2}] we will develop a 
very effective scheme for their calculation. Although for Dirichlet 
and Robin conditions the coefficients are already completely known 
up to $A_2$, we will describe our procedure by starting with these 
low coefficients. We will see that the lower the coefficient the more 
restrictive is the special case, ball calculation. This opens up
for future applications the possibility of applying our approach to 
spectral boundary conditions [\pref{APS,Grubb,GandS1}] and to boundary 
conditions involving tangential derivatives discussed recently in the context 
of the quantization of gauge fields in the presence of boundaries 
[\pref{McandO,AandE,AandE1,AandE2}]. 

In what follows we will take the standpoint that the volume part of 
the coefficients is known. This is motivated by the fact that its
calculation is purely algebraic and very effective schemes already 
exist [\pref{ivan91,fulken88,ven97}]. 
In contrast, the boundary contributions are not 
determined purely algebraically and their evaluation turns out to be
much more involved. It is here that our special case evaluation 
of the smeared coefficients on the ball gives the additional information 
necessary for the complete calculation 
of the coefficients. We will show the effectiveness of the 
scheme by giving all of $A_{5/2}$, but we first explain things 
in detail starting from the lower coefficients.

Some notation is needed.
Here and in the following $F[\cam]=\int_{\cam}dx\, F(x) $
and $F[\partial \cam] = \int_{\partial \cam}\, dy F(y) $,
with $dx$ and $dy$ being the Riemannian volume elements
of $\cam$ and $\partial \cam$. In addition, "$;$" denotes differentiation
with respect to the Levi-Civita connection of $\cam$ and "$:$"
covariant differentiation tangentially with respect to the Levi-Civita
connection of the boundary. Finally, our sign convention is
$R^i_{\phantom{i}jkl} = -\Gamma^i_{jk,l}+
\Gamma^i_{jl,k} +
\Gamma^i_{nk} \Gamma^n_{jl} -
\Gamma^i_{nl} \Gamma^n_{jk}
$ (see for example [\pref{HandE}]).
To state the general form of the coefficients define the partial differential
operator
$$
P= -\Delta - E 
$$
together with Dirichlet or Robin boundary conditions,
$$ 
\cab ^- \phi \equiv \phi|_{\partial \cam } \quad
\mbox{and} \quad \cab_S^+ \phi \equiv \left( \phi_{;m} -S\phi \right)
|_{\partial \cam}   . 
$$
To have a uniform notation we set $S=0$ for Dirichlet boundary conditions
and write $B_S^\mp$. Let $D_\cab$ be the operator defined by the appropriate
boundary conditions. If $F$ is a smooth function on $\cam$, there is an
asymptotic series as $t \to 0$ of the form
$$
\mbox{Tr} _{L^2} \left( F e^{-tP_\cab} \right) \approx
\sum_{n\geq 0 } t^{\frac{n-m} 2 } a_n (F,P_\cab) , 
$$
where the $a_n (F,P_\cab)$ are locally computable [\pref{Gilkey1}]. 

We now state, one by one, the general form of the coefficients and compare
them with our special case evaluation. For convenience we will drop 
the index $\cab$ of the operator $P$. The coefficient $A_0$ is, 
by normalization, 
$$
A_0 (F,P) = (4\pi )^{-D/2} F[\cam ].
$$
The next one is 
$$
A_{1/2} (F,P)  = \delta (4\pi )^{-d/2} F[\partial \cam] .
$$
For the ball this means 
$$
A_{1/2} (F,P)  = \delta (4\pi )^{-d/2} F(1) {\rm vol} (S^d). 
$$
Using the relations (\peq{as3}) and (\peq{adeltafinal})-(\peq{finalshift})
we can immediately determine $\delta$, 
$$
\delta = \left( -\frac 1 4 ^- , \frac 1 4 ^+ \right) .
$$
The coefficient $A_{1/2}$ is thus given for a general manifold from 
the result on the ball (which was clear of course). 
Passing on to $A_1$, the general form is
$$
A_1( F,P) = (4\pi )^{-D/2} 6^{-1} \left\{
 ( 6 FE + FR ) [\cam ] + (b_0 FK +b_1 F_{;m} + b_2 F S ) [\partial \cam]
\right\} 
$$
In our special case on the ball, $K_a^b = \delta_a^b$ and thus 
$$
A_1(F,P) = (4\pi )^{-D/2} 6^{-1} {\rm vol} (S^d)  \left\{
 b_0 F(1) d + b_1 F' (1) + b_2 F(1) S \right\}.
$$
Comparing with the results given in the previous section one finds 
$$
b_0 = 2 ; \qquad b_1 = (3^-, -3^+ ); \qquad b_2 = 12 .
$$
Thus our special case also gives the entire $A_1$ coefficient without 
any further information being needed. 
It is very important that the calculation can be performed 
for an arbitrary ball dimension, $D$, and also for a 
smearing function $F(r)$. This allows one just to compare polynomials 
in $d$ with the associated extrinsic curvature terms in the general 
expression and simply to read off the universal constants in this expression. 

The idea is now clear and in the following
we will state only the general expression and the restrictions 
found from the special case presented in the previous section. We
continue with the next higher coefficient, with the general form,
$$\eqalign{
A_{3/2} (F,P) & = {\delta\over96 (4\pi )^{d/2}} \bigg(\!
     F\big(c_0 E +c_1 R\! +c_2 R_{mm} +c_3 K^2\! +c_4 K_{ab} K^{ab} 
        c_7 SK\! +c_8 S^2\!\big)\cr
&\phantom{\delta (4\pi )^{-d/2} 96^{-1}}
            +F_{;m} (c_5 K +c_9 S) +c_6 F_{;mm} \bigg) .\cr}
$$
The ball calculation immediately gives $7$ of the $10$ unknowns,
$$\eqalign{
&c_3 = (7^-,13^+), \quad  c_4 = (-10 /, 2^+), \quad  c_5 = (30^-, -6^+), \cr 
& c_6 = 24, \quad
c_7 = 96, \quad   c_8 = 192, \quad  c_9 = -96 \cr}.
$$

We next apply the lemma on product manifolds mentioned above [\pref{BandG2}].
Let $N^{\nu} (F) = F_{;m...}$ be the $\nu^{th}$ normal covariant
derivative.
There exist local
formulae $a_n (x,P)$ and $a_{n,\nu} (y,D)$ so that
$$
A_n (F,P,\cab _S^\mp ) = \{FA_n (x,P)\} [\cam] +
 \{  \sum_{\nu = 0}^{2n-1} N^\nu (F) A_{n,\nu} (y,P,\cab_S^\mp )\}
   [\partial \cam].
$$
Let $\cam = \cam _1 \times \cam _2$ and $P=P_1 \otimes 1 + 1\otimes P_2$
and $\partial \cam _2  = \emptyset $. Then
$$
A_{n,\nu} (y,P,\cab _S ^\mp ) = \sum_{p+q=n} A_{p,\nu} (y_1,P_1,\cab _S^\mp )
A_q (x_2,P_2).
$$
For $A_{3/2}$ this means 
$$
A_{3/2} (y,P) = A_{3/2} (y_1,P_1) A_0 ( x_2, P_2) +A_{1/2} (y_1,P_1) A_1 
(x_2, P_2).
$$
We will choose $P_1 = -\Delta_1$ and $P_2 = -\Delta_2 + E (x_2)$ with 
obvious notation to obtain 
$$
\delta 96^{-1} (c_0 E + c_1 R(\cam _2) ) = \delta 6^{-1} (6 E + R (\cam _2))
$$
where we used in addition $R( \cam_1 \times \cam_2 ) = R(\cam _1) +
R(\cam _2 )$. 
This gives 
$$ 
c_0 = 96, \quad c_1 = 16.
$$
It is seen, that the determination of $A_{3/2}$ is relatively simple, once
the ball result is to hand. The lemma on product manifolds is also 
very easily applied and already only one of the universal constants $c_i$, 
namely $c_2$, is missing. 

The remaining information is obtained using the relations between the heat-
kernel coefficients under conformal rescaling [\pref{BandG2}],
$$ 
\frac d {d\ep} |_{\ep =0} a_n \left( 1, e^{-2\ep F} P\right))-
(D-2n) a_n (F,P) = 0
\eql{conf1}
$$
Setting to zero the coefficients of all terms in (\peq{conf1}) gives several
relations between the universal constant $c_i$. We will need only one of 
them. Thus, setting to zero the coefficient of $F_{;mm}$ gives
$$
\frac 1 2 (D-2) c_0 - 2 (D-1) c_1 -(D-1) c_2 -(D-3) c_6 =0 
$$
and so $c_2 = -8$ for Dirichlet and Robin boundary conditions. 
This completes the calculation of $A_{3/2}$. 

We continue with the treatment of $A_2$. Its general form is [\pref{BandG2}],
$$\eqalign{
A_2 (F,P) &= (4\pi )^{-D/2} 360 ^{-1} \left\{
F( 60 \Delta E +60 RE +180 E^2 +12 \Delta R +5 R^2 -2 R_{ij} R^{ij} \right. \cr
       &    +2 R_{ijkl} R^{ijkl} ) [\cam ] 
+F( d_1 E_{;m}+d_2  R_{;m} + d_3 K_{:a}^{a} +d_4 K_{ab:}^{\phantom{ab:}ab}
+d_5 EK \cr 
&+d_6 RK +d_7 R_{mm} K  
    + d_8R_{ambm} K^{ab} + d_9 R_{abc}^{\phantom{abc}b} K^{ac} +
   d_{10} K^3 \cr
& +d_{11} K_{ab} K^{ab} K +d_{12} K_{ab} K^b_c K^{ac} 
 +d_{13} SE + d_{14} SR +d_{15} SR_{mm} \cr
& +d_{16} SK^2 +
  d_{17} S K_{ab} K^{ab} +d_{18} S^2 K +d_{19} S^3 +d_{20} S_{:a}^a) \cr
& +F_{;m} (e_1 E +e_2 R +e_3 R_{mm} +e_4 K^2 +e_5 K_{ab} K^{ab}
+e_8 SK +e_9 S^2 \cr
& \left. +F_{;mm} (e_6 K +e_{10} S) +e_7 (\Delta F)_{;m} \right\} \cr}
\eql{a2gen}
$$
The ball calculation gives
$$\eqalign{
& d_{10} = (40/21^-,40/3^+), \quad d_{11} = (-88/7^-,8^+), \quad 
  d_{12} = (320/21^-,32/3 ^+), \cr
& d_{16} = 144, \quad d_{17} = 48, \quad d_{18} = 480, \quad d_{19} = 480 \cr
& e_4 = (180/7^-,-12^+), \quad e_5 =(-60/7^-,-12^+), \quad 
 e_6 = 24, \cr
& e_7 = (30^-, -30^+), 
\quad   e_8 = -72, \quad e_9 = -240, \quad e_{10} = 120. \cr}
\eql{a2ball}
$$ 
The product formula here reads
$$\eqalign{
A_2 (y,P) = A_2 (y_1,P_1) A_0 (x_2,P_2)& -A_0 (y_1,P_1) A_2 (x_2,P_2)\cr 
         & + A_1 (y_1,P_1) A_1 ( x_2,P_2) \cr}
$$
and leads to the universal constants,
$$\eqalign{
& d_5 = 120, \quad d_6 = 20, \quad d_{13} = 720, \quad d_{14} = 120 \cr
& e_1 = (180^-, -180 ^+), \quad e_2 = (30^-, -30^+). \cr}
\eql{a2pro}
$$
These two inputs already give $20$ of the $30$ unknowns, the 
remaining $10$ are determined by the conformal rescaling 
(\peq{conf1}),
$$   
\frac d {d\ep} |_{\ep =0} a_2 \left( 1, e^{-2\ep F} P\right))-
(D-4 ) a_2 (F,P) = 0.
\eql{confa2}
$$
Having already evaluated many of the constants only a few more relations are
required to fix the remaining ones. In the following list, 
we give, on the left, the term in (\peq{confa2}) whose coefficient 
is equated to zero.
$$\eqalign{
\underline{\mbox{Term}} & \hspace{1cm}\underline{\mbox{Coefficient}} \cr
EF_{;m} & \hspace{1cm} 0 = -2d_1 +60 (D-6) +d_5 (D-1) 
                  -(D-4) e_1 -\frac 1 2 (D-2) 
               d_{13} \cr
(\Delta F)_{;m} & \hspace{1cm} 0= 6 (D-6) +\frac 1 2 (D-2) d_1 
                 -2(D-1) d_2 -(D-4) e_7 \cr
F_{:a} K_:^a & \hspace{1cm} 0= -4 (D-6) +(D-4) d_3 -\frac 1 2 (D-2) d_5 
             +2(D-1) d_6 +d_7 +d_9 \cr
KF_{;mm} & \hspace{1cm} 0= \frac 1 2 (D-2) d_5 -2 (D-1) d_6 -(D-1) d_7 
            -d_8 -(D-4) e_6 \cr
K_{ab:} ^{\phantom{ab:}b} F_:^a & \hspace{1cm} 0= (D-4) d_4 +d_8 
           +(D-3) d_9 +4(D-6) \cr
R_{mm} F_{;m} & \hspace{1cm} 0= (D-1) d_7 +d_8 -2d_9 +e_3 +
                               4(D-6) -\frac 1 2 (D-2) d_{15} \cr
F_{:a} S_:^a & \hspace{1cm} 0= -\frac 1 2 (D-2) d_{13} +2(D-1) d_{14} 
             +d_{15} +(D-4) d_{20} \cr}
\eql{a2conf}
$$
From here one finds the universal constants 
$$\eqalign{
& d_1 = (120^-,-24-^+), \quad d_2 = (18^-,-42^+), \quad d_3 = 24, \quad d_4=0
\cr
& d_7=-4, \quad d_8 = 12, \quad d_9 = -4, \quad d_15 = 0, \quad 
d_{20} = 120, \quad e_3 = 0. \cr} 
\eql{a2con}
$$
This completes the evaluation of $A_2$ and we finally come to the calculation 
of $A_{5/2}$ which, for an arbitrary smearing function $F$, has been
calculated only for a totally geodesic boundary $\partial\cam$.
When $F=1$, it has been determined for $\cam$ a domain of $\oR ^m$. 

It has been shown that for a smooth, but not necessarily totally geodesic,
boundary there exist universal constants such that
$$\eqalign{
A_{5/2} & (F, P) = \mp 5760^{-1}(4\pi)^{-(m-1)/2}
\{F\left\{g_1 E_{;mm} +g_2 E_{;m} S + g_3 E^2 \right. \cr
&+g_4  E_{:a}^{\phantom{:a}a}
+ g_5 RE + 120 \Omega_{ab} \Omega^{ab} + g_6 \Delta R +
g_7 R^2 +g_8 R_{ij} R^{ij} +g_9  R_{ijkl} R^{ijkl}\cr
& +g_{10} R_{mm} E
+g_{11} R_{mm} R +g_{12} RS^2 +(-360^-,90^+)\Omega_{am}
\Omega^a_{\phantom{a} m} +g_{13} R_{;mm}
\cr
&  +g_{14} R_{mm:a}^{\phantom{mm:a}a} 
  +g_{15} R_{mm;mm}
+g_{16} R_{;m} S+g_{17} R_{mm} S^2
+g_{18} S S_{:a}^{\phantom{:a}a} +g_{19} S_{:a} S_:^a
\cr}
$$
$$\eqalign{ 
& +g_{20} R_{ammb} R^{ab} 
  \left.+g_{21} R_{mm} R_{mm}
+g_{22} R_{ammb} R^{a\phantom{mm}b}_{\phantom{a}mm} +
g_{23} ES^2 +g_{24} S^4\right\}  \cr
&   +F_{;m} \left\{g_{25} R_{;m} +g_{26}RS +g_{27} R_{mm} S
+g_{28} S_{:a} ^{\phantom{:a}a}
+g_{29} E_{;m} +g_{30} ES +g_{31} S^3\right\}   \cr
&   +F_{;mm} \left\{g_{32} R +g_{33} R_{mm} +g_{34} E +g_{35} S^2 \right\}
+g_{36}  S F_{;mmm} +g_{37} F_{;mmmm}  \cr
&  +F\left\{d_1 KE_{;m} +d_2 KR_{;m} +
d_3 K^{ab} R_{ammb;m} +d_4 K S_{:b}^{\phantom{:b}b}
+d_5 K_{ab} S_:^{ab} \right.  \cr
&   +d_6 K_{:b} S_:^b +d_7 {K_{ab:}}^a S_:^b
+d_8 {K_{:b}}^b S +d_9 {K_{ab:}}^{ab} S
+ d_{10} K_{:b} K_:^b
+d_{11} {K_{ab:}}^a K_:^b   \cr
&   +d_{12} {K_{ab:}}^a {K^{bc}}_{:c}
+d_{13} K_{ab:c} K^{ab\phantom{:}c}_{\phantom{ab}:}
+d_{14} K_{ab:c} K^{ac\phantom{:}b}_{\phantom{ac}:}
+d_{15} K_{:b}^{\phantom{:b}b} K  \cr
&   +d_{16} K_{ab:}^{\phantom{ab:}ab} K
+d_{17} K_{ab:\phantom{a}c}^{\phantom{ab:}a} K^{bc}
+d_{18} K_{:bc}K^{bc} +d_{19} K_{bc:a}^{\phantom{bc:a}a}
K^{bc}   \cr
&   +g_{38} KSE +d_{20} KS R_{mm} +g_{39} KSR
+d_{21} K_{ab} R^{ab} S +d_{22} K^{ab} S R_{ammb}
\cr
& +g_{40} K^2 E  
  +g_{41} K_{ab} K^{ab} E
+g_{42} K^2 R
+g_{43} K_{ab}K^{ab} R +d_{23} K^2 R_{mm}   \cr
&   +d_{24} K_{ab} K^{ab} R_{mm} +d_{25} KK_{ab} R^{ab}
+d_{26} KK^{ab} R_{ammb} +d_{27}K_{ab}K^{ac}R^b_c   \cr
&   +d_{28} K_a^b K^{ac} R_{bmmc} +d_{29} K_{ab} K_{cd} R^{acbd} +
d_{30} KS^3 +d_{31} K^2 S^2 +d_{32} K_{ab} K^{ab} S^2  \cr
&   +d_{33} K^3 S
+d_{34} KK_{ab} K^{ab} S +d_{35} K_{ab} K^{bc} K^a_c S +
d_{36} K^4 +d_{37} K^2 K_{ab} K^{ab}  \cr
&   \left.+d_{38} K_{ab}K^{ab} K_{cd} K^{cd} + d_{39}
KK_{ab} K^{bc} K_c^a + d_{40} K_{ab} K^{bc} K_{cd} K^{da} \right\}
       \cr
&   +F_{;m} \left\{g_{44} KE +d_{41} KR_{mm} +
g_{45} KR+d_{42} KS^2
\right.    \cr
&  +d_{43} K_{:b}^{\phantom{:b}b} +
d_{44} K_{ab:}^{\phantom{ab:}ab}
+d_{45} K_{ab}R^{ab} +d_{46} K^{ab} R_{ammb}
+d_{47} K^2S \cr
&  \left.+d_{48}K_{ab} K^{ab} S +d_{49} K^3
+d_{50} KK_{ab} K^{ab} +d_{51} K_{ab} K^{bc} K^a_c
\right\}   \cr
&    +F_{;mm} \left\{d_{52} KS +d_{53} K^2 +d_{54}
K_{ab} K^{ab} \right\} +d_{55} K  F_{;mmm}\} [\partial {\cal M}]
\cr}
\eql{4}
$$
In this case specializing to the ball gives,
$$\eqalign{
g_{24} = 1440 & \hspace{1cm} g_{31} = -720 \cr
g_{35} = 360 & \hspace{1cm}
g_{36} = -180 \cr
g_{37} = 45 & \hspace{1cm} d_{30} = 2160 \cr
d_{31} = 1080 & \hspace{1cm}
d_{32} = 360 \cr
d_{33} = 885/4 & \hspace{1cm}
 d_{34} = 315/2  \cr}
$$
$$\eqalign{
d_{35} = 150 & \hspace{1cm}  d_{36} = (-65/128^-, 2041/128^+)   \cr
d_{37} = (-141/32^-,417/32^+) & \hspace{1cm}
d_{40} = (-327/8^-, 231/8^+)  \cr
d_{42} = -600 & \hspace{1cm}
 d_{47} = -705/4   \cr
d_{48} = 75/2 & \hspace{1cm} d_{49}
= (495/32^-, -459/32^+)  \cr
 d_{50} = (-1485/16^-, -267/16^+) & \hspace{1cm}
d_{51} = (225/2^-, 54^+)   \cr
d_{52} = 30 & \hspace{1cm}
d_{53} = (1215/16^-, 315/16^+)   \cr
d_{54} = (-945/8^-, -645/8^+) & \hspace{1cm}
d_{55} = (105^-,30^+)
  \cr}
$$
and $d_{38} +d_{39} =(1049/32^-, 1175/32^+)$.

The product formula explicitly reads
$$\eqalign{
A_{5/2} (y,P ) = A_{5/2} (y_1,P_1 ) A_0 (x_2,P_2)+&
A_{3/2} (y_1,P_1 ) A_1 (x_2,P_2) \cr
&+A_{1/2} (y_1,P_1 )A_2 (x_2,P_2), \cr}
$$
which gives the $22$ universal constants,
$$\eqalign{
& g_3 = 720, \quad g_5 = 240,  \quad g_6 = 48, \quad  g_7 = 20, \cr
& g_8= -8, \quad g_9 = 8, \quad g_{10} = -120, \quad g_{11}=-20, \cr
&  g_{12} = 480, \quad   g_{23} = 2880, \quad g_{26} = -240, \quad 
g_{30} =-1440, \cr
& g_{32} = 60, \quad g_{34} = 360, \quad g_{38} = 1440, \quad g_{39} = 240\cr
& g_{40} = (105^-, 195^+), \quad g_{41} = (-150^-, 30^+), \quad
g_{42} =
(105/6^-, 195/6^+) \cr
&  g_{43} = (-25^-,5^+), \quad 
  g_{44}= (450^-,-90^+), \quad g_{45} = (75^-,-15^+). \cr}
$$

All this information puts us in a very good position to use the 
relations between the heat kernel coefficients that result from conformal 
rescalings. The relevant relation reads
$$
\frac d {d\ep} |_{\ep =0} A_{5/2} \left( 1, e^{-2\ep F} P\right))-
(D-5) A_{5/2} (F,P) = 0
\eql{8}
$$
Setting to zero the coefficients of all terms in (\peq{8})
we obtain the equations given in 
(\peq{app1}). 
(They are ordered in such a way that
nearly every equation immediately yields a universal constant. This
was the main motivation for the given ordering.)
\begin{ignore}
$$\eqalign{
\underline{\mbox{Term}} & \hspace{1cm} \underline{\mbox{Coefficient}} \cr
EF_{;mm} & \hspace{1cm} 0=-2g_1 +(D-2)g_3 -2(D-1)g_5 
                             -(D-1)g_{10} - (D-5) g_{34} \cr
ESF_{;m} &  \hspace{1cm}  0=-2g_2 -(D-2)g_{23} +(D-1) g_{38} -(D-5) g_{30} \cr
SF_{;mmm} & \hspace{1cm} 0=\frac 1 2 (D-2) g_2 -2(D-1)g_{16} -(D-5) g_{36} \cr
KSF_{;mm} & \hspace{1cm} 0=\frac 1 2 (D-2) g_2 -2(D-1) g_{16} 
+\frac 1 2 (D-2) g_{38}
         -(D-1) d_{20} \cr
  & \hspace{1cm}\phantom{0=}-2(D-1) g_{39} -d_{21} +d_{22} -(D-5) d_{52} \cr
FE_{:a}^{\phantom{:a} a} &\hspace{1cm}  0=-g_1 +(D-2) g_3 -(D-5) g_4 -2(D-1) 
g_5 -g_{10} \cr
F_{;mmmm} & \hspace{1cm} 0=\frac 1 2 (D-2) g_1 -2(D-1) g_6 -2(D-1) g_{13} 
-(D-1) g_{15}
            -(D-5) g_{37}  \cr
F\Delta R & \hspace{1cm} 0=\frac 1 2 (D-2) g_5 -(D-4) g_6 -4(D-1) g_7 -Dg_8 
-4g_9 -g_{11}
           -g_{13} +\frac 1 2 g_{20} \cr
FR_{;mm} & \hspace{1cm}  0= -\frac 1 2 (D-2) g_5 +(D-4) g_6 +4(D-1) g_7 
+2(D-1) g_8 +8g_9
            +g_{11} \cr
&   \hspace{1cm} \phantom{0=}+g_{13} -2g_{15} -\frac D 2 g_{20} +g_{22} \cr
FR_{mm:a}^{\phantom{mm:a} a} &\hspace{1cm} 0=\frac 1 2 (D-2) g_1 -2(D-1) 
g_6 +\frac 1 2
       (D-2) g_{10} -2(D-1) g_{11} -2(D-1) g_{13} \cr
 &\hspace{1cm}  \phantom{0=} -(D-5) g_{14} -2g_{15}
                +(D-1)
        g_{20} -2g_{21} -2g_{22}  \cr
F_{;mm} S^2 & \hspace{1cm} 0=-2(D-1) g_{12} -(D-1) g_{17} +\frac 1 2 (D-2) 
g_{23}
            -(D-5) g_{35}  \cr
FS_{:a} S_:^a & \hspace{1cm} 0= -4(D-1) g_{12} -2g_{17} -(D-3) g_{18} +2g_{19}
                +(D-2) g_{23}  \cr
F_{;m} E_{;m} & \hspace{1cm} 0= -5g_1 -\frac 1 2 (D-2) g_2 +(D-1) d_1 -(D-5) 
g_{29} \cr
F_{;mmm} K & \hspace{1cm} 0= \frac 1 2 (D-2) g_1 -4(D-1) g_6 -2(D-1) g_{13} 
-g_{15}
        +\frac 1 2 (D-2) d_1   \cr
 &\hspace{1cm}  \phantom{0=} -2(D-1) d_2 +d_3 -(D-5) d_{55} \cr
F_{;m} R_{;m} &\hspace{1cm}  0= -\frac 1 4 (D-2) g_1 +(2D-7) g_6 +(D-6) 
g_{13} -2g_{15}\cr
&\hspace{1cm}  \phantom{0=}
          -\frac 1 2 (D-2) g_{16} +(D-1) d_2 -\frac 1 2 d_3 -(D-5) g_{25}\cr
F_{;mm} R_{mm} &\hspace{1cm}  0= -(D-2) g_1 +4(D-1) g_6 -2(D-2) g_8 -8g_9 
+\frac 1 2
          (D-2) g_{10} \cr
& \hspace{1cm}  \phantom{0=}  -2(D-1) g_{11}
              +4(D-1) g_{13} -2(D-1) g_{21} -2g_{22}
               -(D-5) g_{33}   \cr
F_{;m} R_{mm} S &\hspace{1cm}  0= -\frac 1 2 (D-2) g_2 +2(D-1) g_{16} -(D-2) 
g_{17}
          +(D-1) d_{20}  \cr
& \hspace{1cm} \phantom{0=} -d_{21} -d_{22} -(D-5) g_{27}   \cr
FKS_{:a}^{\phantom{:a} a} & \hspace{1cm} 0= -(D-4) d_4 -d_5 +d_6 +d_7 -d_8 
-d_9 +\frac
             1 2 (D-2) g_{38} \cr}
$$
$$\eqalign{
 &  \hspace{1cm} \phantom{0=}-d_{20} -2(D-1) g_{39} -d_{21}  \cr
FK_{:a}^{\phantom{:a} a} S & \hspace{1cm}= -\frac 1 2 (D-2) g_2 +2(D-1) g_{16}
                   -d_4 +d_6 -(D-4) d_8  \cr
 &\hspace{1cm}  \phantom{0=} +\frac 1 2 (D-2) g_{38} -d_{20}
           -2(D-1) g_{39} -d_{21}  \cr
FK_{ab} S_:^{ab} & \hspace{1cm} 0= -(D-2) g_2 +4(D-1) g_{16} +3d_5 -(D-2) d_7
                +(D-2) d_9  \cr
  &\hspace{1cm}  \phantom{0=} -(D-2) d_{21} +d_{22}  \cr
FK_{ab:}^{\phantom{ab:} ab} S &\hspace{1cm}  0= -d_5 +d_7 -(D-4) d_9 -(D-2) 
d_{21} +d_{22}\cr
F_{;m} S_{:a} ^{\phantom{:a} a} & \hspace{1cm} 0= \frac 1 2 (D-2) g_2 -2(D-1) 
g_{16} -(D-2)
             g_{18} +(D-2) g_{19} +(D-1) d_4 \cr
& \hspace{1cm} \phantom{0=}  +d_5
                      -(D-1) d_6 -d_7
             +(D-1) d_8 +d_9 -(D-5) g_{28}
\cr}
$$
\end{ignore}
Using the relation (\peq{app1})
we find
$$\eqalign{
& g_1 = 360, \quad g_2 = -1440, \quad  g_4 = 240, \cr
& g_{13} = 12, \quad g_{14} = 24, \quad g_{15} = 15, \cr
& g_{16} = -270, \quad  g_{17} = 120, \quad g_{18} = 960, \cr
& g_{19} = 600, \quad g_{20} = -16, \quad  g_{21} = -17, \cr
& g_{22} = -10 , \quad g_{25} = (60^-,195/2^+), \quad g_{27} = 90, \cr
& g_{28} = -270, \quad g_{29} = (450^-,630^+), \quad g_{33} = -90, \cr
& d_1 = (450^-,-90^+), \quad d_2 = (42^-,-111/2^+), \quad d_3 = (0^-,30^+)\cr
& d_4 = 240, \quad d_5 = 420, \quad  d_6 = 390, \cr
& d_7 = 480 , \quad d_8 = 420, \quad d_9 = 60, \cr
& d_{20} = 30, \quad d_{21} = -60 , \quad  d_{22} = -180
\cr}
$$
Thus the equations given up to this point allow for the determination
of the universal constants apart from two groups. The first group
is $d_{23},...,d_{29},d_{38},d_{39}, d_{41}$, $d_{45}, d_{46}$ and
the second one, $d_{10},...,d_{19}, d_{43}, d_{44}$.
The first group is completely determined using the relations
given in (\peq{app2}).
\begin{ignore}
$$\eqalign{
\underline{\mbox{Term}} & \hspace{1cm} \underline{\mbox{Coefficient}}\cr
F_{;mm} K_{ab} K^{ab} & \hspace{1cm} 0= -(D-2) g_1 +4(D-1) g_6 +4(D-1) g_{13}
          +2g_{15} +d_3 +\frac 1 2 (D-2) g_{41}\cr
 & \hspace{1cm} \phantom{0=}  -2(D-1) g_{43} -(D-1) d_{24}
          -d_{27} +d_{28} -(D-5) d_{54}\cr
F_{;mm} K^2 & \hspace{1cm} 0= -2(D-1) g_6 +\frac 1 2 (D-2) d_1 -2(D-1) d_2 +
              (D-2) g_{40}\cr
& \hspace{1cm} \phantom{0=} -2(D-1) g_{42}
                -(D-1) d_{23} -d_{25} +d_{26}
                -(D-5) d_{53}\cr
F_{;m} KR & \hspace{1cm} 0= \frac 1 2 (D-2) g_5 -2g_6 -4(D-1) g_7 -2g_8 
-g_{11} -2d_2 -\frac
        1 2 (m-2) g_{39} \cr
 &\hspace{1cm}  \phantom{0=} +2(D-1) g_{42} +2g_{43} +d_{25} -(D-5) g_{45}\cr}
$$
$$\eqalign{
F_{;m} KR_{mm} &\hspace{1cm}  0= \frac 1 2 (D-2) g_1 +\frac 1 2 (D-2) g_{10}
         -2(D-1) g_{11} -2(D-1) g_{13} +4g_{15}\cr
  & \hspace{1cm} \phantom{0=} +g_{20}
           -2g_{21} -\frac 1 2
         (D-2) d_1 +2(m-1) d_2 +d_3 -\frac 1 2 (D-2) d_{20}\cr
  & \hspace{1cm} \phantom{0=} +2(D-1) d_{23}
           +2d_{24} -d_{25} -d_{26} -(D-5) d_{41}\cr
F_{;m} K_{ab} R^{ab} & \hspace{1cm} 0= -\frac 1 2 (D-2) g_1 +2(D-1) g_6 
-2(D-2) g_8 -8g_9
        +2(D-1) g_{13} \cr
   & \hspace{1cm} \phantom{0=} -4g_{15} +g_{20}
          -d_3 -\frac 1 2 (D-2) d_{21}
          +(D-1) d_{25} +2d_{27}\cr
   &\hspace{1cm}  \phantom{0=}  +2d_{29} -(D-5) d_{45} \cr
F_{;m} K^{ab} R_{ammb} & \hspace{1cm} 0= -(D-2) g_1 +4(D-1) g_6 +4(D-1) g_{13}
          +2g_{15} -(D-2) g_{20}\cr
   & \hspace{1cm} \phantom{0=} +2g_{22} -d_3
         -\frac 1 2 (D-2) d_{22}
          +(D-1) d_{26} +2d_{28}\cr
  &\hspace{1cm}  \phantom{0=}  +2d_{29} -(D-5) d_{46}\cr
F_{;m} K_{ab} K^{bc} K_c^a & \hspace{1cm} 0= (D-2) g_1 -4(D-1) g_6 -4(D-1) 
g_{13}
          -2g_{15} -d_3 -(D-2) d_{27}\cr
 & \hspace{1cm} \phantom{0=}  +d_{28}
            +2d_{29}
          -\frac 1 2 (D-2) d_{35} +(D-1) d_{39} +4d_{40} -(D-5) d_{51}\cr
FR_{ac} K^c_b K^{ab} & \hspace{1cm} 0= -2(D-2) g_8 -8g_9 +4g_{15} +g_{20} +2d_3
           +4d_{13} +4d_{14} -4d_{19}\cr
 & \hspace{1cm} \phantom{0=}  -(D-2) d_{27} +d_{28} +2d_{29}
\cr}
$$
\end{ignore}
One finds
$$\eqalign{
& d_{23} = (-215/16^-,-275/16^+), \quad d_{24} = (-215/8 ^-, -275/8^+),\cr
& d_{25} = (14^-,-1^+) , \quad d_{26} = (-49/4^-, -109/4^+) ,\cr
& d_{27} = 16, \quad d_{28} = (47/2^-,-133/2^+), \cr
& d_{29} = 32, \quad  d_{38} = (777/32^-, 375/32^+) , \cr  
& d_{39} = (17/2^-, 25^+), \quad d_{41} = (-255/8^-,165/8^+), \cr
&  d_{45} =(-30^-,-15^+), \quad d_{46} = (-465/4^-,-165/4^+)
\cr}
$$
Finally we consider the second group mentioned above.
As we will see, one needs just one more relation in addition to those
obtained from equation (\peq{8}), which are presented in (\peq{app3}).
\begin{ignore}
$$\eqalign{
\underline{\mbox{Term}} & \hspace{1cm} \underline{\mbox{Coefficient}}\cr
FK_{:b}K^{:b} & \hspace{1cm} 0=2(D-1) g_6 -4g_{15} -(D-2) g_{20} +2g_{22}
             - \frac 1 2 (D-2) d_1 \cr
   &\hspace{1cm}  \phantom{0=} +2(D-1) d_2 +2d_{10}
              +d_{11}
               -(D-3) d_{15} -d_{16} -d_{18} +(D-2) g_{40}\cr
  &\hspace{1cm}  \phantom{0=}
              -4(D-1) g_{42} -2d_{23} -2d_{25}\cr}
$$
$$\eqalign{
FK_{ab:}^{\phantom{ab:}a} K_:^b &\hspace{1cm}  0=2(D-2) g_1 -4(D-1) g_6 
-8(D-1) g_{13}
                +(4D-6) g_{20} -8g_{22}\cr
 & \hspace{1cm} \phantom{0=}  -(D-2) d_1
        +4(m-1) d_2 -(D-3) d_{11}
                +2d_{12} -2d_{14} +2d_{16}\cr
  & \hspace{1cm} \phantom{0=} -2d_{17} +2d_{18} -2(D-2) d_{25}
                +2d_{26} -4d_{29} \cr
FK_{ab:c} K^{ab\phantom{:}c}_{\phantom{ab}:} & \hspace{1cm} 0=(D-2) g_1 
-4(D-1) g_6
              - 4(D-1) g_{13} -2g_{15} +(D-2) g_{20}\cr
  &\hspace{1cm}  \phantom{0=}  -2g_{22} -3d_3
              +2d_{13} +2d_{14} -(D-3) d_{19} +(D-2) g_{41} \cr
&\hspace{1cm}  \phantom{0=}
               -4(D-1) g_{43} -2d_{24} -2d_{27}  \cr
FKK_{ab:}^{\phantom{ab:}ab} & \hspace{1cm} 0= 4(D-2) g_8 +16g_9 -4g _{15} 
-Dg_{20} +2g_{22}
                +d_{11} +2d_{12} -(D-4) d_{16}\cr
          &\hspace{1cm}  \phantom{0=} -2d_{17} -d_{18} -(D-2) d_{25}
                +d_{26} -2d_{29}\cr
F_{;m} K_{:a}^{\phantom{:a} a} &\hspace{1cm}  0= -\frac 3 2 (D-2) g_1 
+4(D-1) g_6 -4(D-2) g_8
                -16g_9 +6(D-1) g_{13}\cr
  & \hspace{1cm} \phantom{0=}  +\frac 1 2 (D-2) d_1
          -2(D-1) d_2 -d_3
                -\frac 1 2 (D-2) d_4 +\frac 1 2 (D-2) d_6\cr
   & \hspace{1cm} \phantom{0=}-\frac 1 2 (D-2)
                 d_8 -2(D-1) d_{10}\cr
       & \hspace{1cm} \phantom{0=}  -d_{11} -2d_{13} +2(D-1) d_{15}
                 +d_{16} +d_{18} +2d_{19} -(D-5) d_{43}\cr
F_{;m} K_{ab:}^{\phantom{ab:}ab} &\hspace{1cm}  0= \frac 1 2 (D-2) g_1 
-2(D-1) g_6
             +4(D-2) g_8 +16g_9 -2(D-1) g_{13}\cr
   & \hspace{1cm} \phantom{0=}  +2g_{15} +2d_3
       -\frac 1 2
            (D-2) d_5 +\frac 1 2 (D-2) d_7 -\frac 1 2 (D-2) d_9 \cr
  & \hspace{1cm} \phantom{0=}
            -(D-1) d_{11} -2d_{12} -2d_{14}
          +(D-1) d_{16} +2d_{17} +
               (D-1) d_{18}\cr
    &\hspace{1cm}  \phantom{0=}-(D-5) d_{44}\cr
FK_{ab:}^{\phantom{ab:} a} K^{bc}_{\phantom{bc}:c} & 
\hspace{1cm} 0= (4-3D) g_{20}
                +6g_{22} -2d_3 -2(D-2) d_{12} -4d_{13} -2d_{14}\cr
    & \hspace{1cm} \phantom{0=}
          +(D+1) d_{17} +4d_{19} -(D-2) d_{27} +d_{28} +2d_{29}\cr
F K_{:ab} K^{ab} & \hspace{1cm} 0= 2(D-2) g_1 -4(D-1) g_6 -2(D-2) g_8 -8g_9
             -8 (D-1) g_{13} \cr
   & \hspace{1cm} \phantom{0=}+(4D-5) g_{20}
           -8g_{22} -(D-2) d_1
             +4(D-1) d_2 -(D-2) d_{11}\cr
   & \hspace{1cm} \phantom{0=} -2d_{14} +(D-2) d_{16}
            +3d_{18}
       -(D-2) d_{25} +d_{26} -2d_{29}
\cr}
$$
\end{ignore}
They yield
$$\eqalign{
& d_{11} = (58^-,238^+), \quad  d_{15} = (6^-, 111^+) , \cr
& d_{16} = (-30^-,-15^+), \quad d_{19} = (54^-,114^+), \cr}
$$
together with the relations
$$\eqalign{
2d_{10} +d_{43} = -91, &\hspace{1cm} 2d_{10} -d_{18} 
         = (-983/8^-,-1403/8^+),\cr
2d_{14} -3d_{18} = (-913/4^-,-2533/4^+), & \hspace{1cm} 
d_{13} +d_{14} = (297/8^-,837/8^+),\cr
d_{18}-d_{44} = (60^-,225^+), &
\hspace{1cm}  2d_{12} -2d_{17} -d_{18} = (-7/4^-,-787/4^+),\cr
2d_{12} -d_{17} = 32. &
\cr}
$$
This is all we can get with the relation (\peq{8}). It is seen that,
given $d_{43}$ or $d_{44}$, for example, the remaining constants can be
determined. This is achieved with the equation [\pref{BandG2}]
$$
\frac d {d\ep} |_{\ep =0} A_{5/2} \left( e^{-2\ep f} F, e^{-2\ep f} P
\right) =0
\quad \mbox{for } D=7.
\eql{9}
$$ 
Thus, finally, one gets
$$\eqalign{
d_{10} =(-413/16^-,487/16^+), & \hspace{1cm} d_{12} = (-11/4^-,49/4^+), \cr
d_{13} = (355/8^-,535/8^+), & \hspace{1cm} 
d_{14} = (-29/4^-,151/4^+), \cr
 d_{17} = (-75/2^-,-15/2^+), & \hspace{1cm} 
d_{18} = (285/4^-,945/4^+), \cr
d_{43} = (-315/8^-,-1215/8^+), &\hspace{1cm}  d_{44} = 45/4 ,
\cr}  
$$
which concludes the calculation of the complete $A_{5/2}$ coefficient 
on a smooth manifold with boundary. 
All terms not displayed in the above lists have been used as checks
on the computed universal constants.
\section{\bf 4. Conclusions}
In this article we have developed a technique for the calculation of 
smeared heat-kernel coefficients on the generalized cone. This 
is a generalization of our previous work [\pref{BKD}] where we treated the 
$F=1$ case only. All technical and aesthetic advantages emphasized previously 
are still present for arbitrary $F$. Namely, by restricting attention to 
the ball, and using a function $F$ as general as needed, the 
coefficients can be found as  polynomials in the dimension of the ball. 
This has the advantage that the special case evaluation 
can easily be used to put restrictions on the general form of the 
heat-kernel coefficients. This idea was applied to the coefficients 
$A_0,...,A_{5/2}$ for Dirichlet and Robin boundary conditions. Supplemented 
by a lemma on product manifolds and relations from conformal rescalings, 
we have shown that starting with the results of the special case, treated 
here for the first time, the complete coefficients are obtained very 
effectively.

The method is clearly capable of being applied to other situations. 
An example of even more complexity is the generalized boundary condition 
involving tangential derivatives of the field 
[\pref{McandO,AandE,AandE1,AandE2}]. Up to now, 
for this case, we have applied the technique of special case evaluation only 
to the $4$-dimensional ball [\pref{DK}]. The treatment of the 
$D$-dimensional ball is under consideration with the aim of finding the 
general form of the coefficients using the ideas presented here. 
Another situation of interest is the spectral boundary condition applied
to spinor fields. These conditions are nonlocal and it is 
known that the relations obtained with conformal techniques are not 
sufficient for the determination of the entire coefficient
[\pref{BGG}]. However, supplemented by the ball calculation, it is possible 
to find at least the lower coefficients in this case too. We reserve 
exposition of these extensions for later.

A general word of caution, however. The evaluation of higher and higher 
coefficients
quickly becomes prodigiously complicated, even for just the volume terms,
and there is the danger of it becoming an end in itself. The question
is whether there is any value in displaying an impenetrable profusion of
terms, without some strong motivation. So far as the boundary terms go, we
feel that with $A_{5/2}$ we probably have reached the limit of what can 
sensibly be calculated and displayed. Already the expressions are becoming
unwieldy. Further progress in this area should,
we think, be limited to extending the class of manifolds, say to those
with non-smooth boundaries, and to the consideration of other fields and
boundary conditions as indicated in the preceding paragraph.

\section{\bf Acknowledgments}
We are indebted to Peter Gilkey for providing results on conformal
rescalings. They have served as a very good check of the calculation and
have been of invaluable help.
Furthermore we wish to thank Giampiero Esposito and Michael Bordag
for interesting discussions.

This investigation has been partly supported by the DFG under contract number
BO1112/4-2 and partly by the EPSRC under grant number GR/L75708.
\section{\bf Appendix}
In this appendix we list the relations resulting from the 
conformal property (\peq{8}).

The first group of relations is
$$\eqalign{
\underline{\mbox{Term}} & \hspace{1cm} \underline{\mbox{Coefficient}} \cr
EF_{;mm} & \hspace{1cm} 0=-2g_1 +(D-2)g_3 -2(D-1)g_5 
                             -(D-1)g_{10} - (D-5) g_{34} \cr
ESF_{;m} &  \hspace{1cm}  0=-2g_2 -(D-2)g_{23} +(D-1) g_{38} -(D-5) g_{30} \cr
SF_{;mmm} & \hspace{1cm} 0=\frac 1 2 (D-2) g_2 -2(D-1)g_{16} -(D-5) g_{36} \cr
KSF_{;mm} & \hspace{1cm} 0=\frac 1 2 (D-2) g_2 -2(D-1) g_{16} +\frac 1 2 
(D-2) g_{38}
         -(D-1) d_{20} \cr
  & \hspace{1cm}\phantom{0=}-2(D-1) g_{39} -d_{21} +d_{22} -(D-5) d_{52} \cr
FE_{:a}^{\phantom{:a} a} &\hspace{1cm}  0=-g_1 +(D-2) g_3 -(D-5) g_4 -2(D-1) 
g_5 -g_{10} \cr
F_{;mmmm} & \hspace{1cm} 0=\frac 1 2 (D-2) g_1 -2(D-1) g_6 -2(D-1) g_{13} 
-(D-1) g_{15}
            -(D-5) g_{37}  \cr
F\Delta R & \hspace{1cm} 0=\frac 1 2 (D-2) g_5 -(D-4) g_6 -4(D-1) g_7 -Dg_8 
-4g_9 -g_{11}
           -g_{13} +\frac 1 2 g_{20} \cr
FR_{;mm} & \hspace{1cm}  0= -\frac 1 2 (D-2) g_5 +(D-4) g_6 +4(D-1) g_7 
+2(D-1) g_8 +8g_9
            +g_{11} \cr
&   \hspace{1cm} \phantom{0=}+g_{13} -2g_{15} -\frac D 2 g_{20} +g_{22} \cr
FR_{mm:a}^{\phantom{mm:a} a} &\hspace{1cm} 0=\frac 1 2 (D-2) g_1 -2(D-1) g_6 
+\frac 1 2
       (D-2) g_{10} -2(D-1) g_{11} -2(D-1) g_{13} \cr
 &\hspace{1cm}  \phantom{0=} -(D-5) g_{14} -2g_{15}
                +(D-1)
        g_{20} -2g_{21} -2g_{22}  \cr
F_{;mm} S^2 & \hspace{1cm} 0=-2(D-1) g_{12} -(D-1) g_{17} +\frac 1 2 (D-2) 
g_{23}
            -(D-5) g_{35}  \cr
FS_{:a} S_:^a & \hspace{1cm} 0= -4(D-1) g_{12} -2g_{17} -(D-3) g_{18} +2g_{19}
                +(D-2) g_{23}  \cr
F_{;m} E_{;m} & \hspace{1cm} 0= -5g_1 -\frac 1 2 (D-2) g_2 +(D-1) d_1 -(D-5) 
g_{29} \cr}
$$
$$\eqalign{
F_{;mmm} K & \hspace{1cm} 0= \frac 1 2 (D-2) g_1 -4(D-1) g_6 -2(D-1) g_{13} 
-g_{15}
        +\frac 1 2 (D-2) d_1   \cr
 &\hspace{1cm}  \phantom{0=} -2(D-1) d_2 +d_3 -(D-5) d_{55} \cr
F_{;m} R_{;m} &\hspace{1cm}  0= -\frac 1 4 (D-2) g_1 +(2D-7) g_6 +(D-6) 
g_{13} -2g_{15}\cr
&\hspace{1cm}  \phantom{0=}
          -\frac 1 2 (D-2) g_{16} +(D-1) d_2 -\frac 1 2 d_3 -(D-5) g_{25}\cr
F_{;mm} R_{mm} &\hspace{1cm}  0= -(D-2) g_1 +4(D-1) g_6 -2(D-2) g_8 -8g_9 
+\frac 1 2
          (D-2) g_{10} \cr
& \hspace{1cm}  \phantom{0=}  -2(D-1) g_{11}
              +4(D-1) g_{13} -2(D-1) g_{21} -2g_{22}
               -(D-5) g_{33}   \cr
F_{;m} R_{mm} S &\hspace{1cm}  0= -\frac 1 2 (D-2) g_2 +2(D-1) g_{16} 
-(D-2) g_{17}
          +(D-1) d_{20}  \cr
& \hspace{1cm} \phantom{0=} -d_{21} -d_{22} -(D-5) g_{27}   \cr
FKS_{:a}^{\phantom{:a} a} & \hspace{1cm} 0= -(D-4) d_4 -d_5 +d_6 +d_7 
-d_8 -d_9 +\frac
             1 2 (D-2) g_{38} \cr
 &  \hspace{1cm} \phantom{0=}-d_{20} -2(D-1) g_{39} -d_{21}  \cr
FK_{:a}^{\phantom{:a} a} S & \hspace{1cm}= -\frac 1 2 (D-2) g_2 +2(D-1) g_{16}
                   -d_4 +d_6 -(D-4) d_8  \cr
 &\hspace{1cm}  \phantom{0=} +\frac 1 2 (D-2) g_{38} -d_{20}
           -2(D-1) g_{39} -d_{21}  \cr
FK_{ab} S_:^{ab} & \hspace{1cm} 0= -(D-2) g_2 +4(D-1) g_{16} +3d_5 -(D-2) d_7
                +(D-2) d_9  \cr
  &\hspace{1cm}  \phantom{0=} -(D-2) d_{21} +d_{22}  \cr
FK_{ab:}^{\phantom{ab:} ab} S &\hspace{1cm}  0= -d_5 +d_7 -(D-4) d_9 
-(D-2) d_{21} +d_{22}\cr
F_{;m} S_{:a} ^{\phantom{:a} a} & \hspace{1cm} 0= \frac 1 2 (D-2) g_2 
-2(D-1) g_{16} -(D-2)
             g_{18} +(D-2) g_{19} +(D-1) d_4 \cr
& \hspace{1cm} \phantom{0=}  +d_5
                      -(D-1) d_6 -d_7
             +(D-1) d_8 +d_9 -(D-5) g_{28}
\cr}
\eql{app1}
$$
For the second one we have
$$\eqalign{
\underline{\mbox{Term}} & \hspace{1cm} \underline{\mbox{Coefficient}}\cr
F_{;mm} K_{ab} K^{ab} & \hspace{1cm} 0= -(D-2) g_1 +4(D-1) g_6 +4(D-1) g_{13}
          +2g_{15} +d_3 +\frac 1 2 (D-2) g_{41}\cr
 & \hspace{1cm} \phantom{0=}  -2(D-1) g_{43} -(D-1) d_{24}
          -d_{27} +d_{28} -(D-5) d_{54}\cr
F_{;mm} K^2 & \hspace{1cm} 0= -2(D-1) g_6 +\frac 1 2 (D-2) d_1 -2(D-1) d_2 +
              (D-2) g_{40}\cr
& \hspace{1cm} \phantom{0=} -2(D-1) g_{42}
                -(D-1) d_{23} -d_{25} +d_{26}
                -(D-5) d_{53}\cr
F_{;m} KR & \hspace{1cm} 0= \frac 1 2 (D-2) g_5 -2g_6 -4(D-1) g_7 -2g_8 
-g_{11} -2d_2 -\frac
        1 2 (m-2) g_{39} \cr
 &\hspace{1cm}  \phantom{0=} +2(D-1) g_{42} +2g_{43} +d_{25} -(D-5) g_{45}\cr}
$$
$$\eqalign{
F_{;m} KR_{mm} &\hspace{1cm}  0= \frac 1 2 (D-2) g_1 +\frac 1 2 (D-2) g_{10}
         -2(D-1) g_{11} -2(D-1) g_{13} +4g_{15}\cr
  & \hspace{1cm} \phantom{0=} +g_{20}
           -2g_{21} -\frac 1 2
         (D-2) d_1 +2(m-1) d_2 +d_3 -\frac 1 2 (D-2) d_{20}\cr
  & \hspace{1cm} \phantom{0=} +2(D-1) d_{23}
           +2d_{24} -d_{25} -d_{26} -(D-5) d_{41}\cr
F_{;m} K_{ab} R^{ab} & \hspace{1cm} 0= -\frac 1 2 (D-2) g_1 +2(D-1) g_6 
-2(D-2) g_8 -8g_9
        +2(D-1) g_{13} \cr
   & \hspace{1cm} \phantom{0=} -4g_{15} +g_{20}
          -d_3 -\frac 1 2 (D-2) d_{21}
          +(D-1) d_{25} +2d_{27}\cr
   &\hspace{1cm}  \phantom{0=}  +2d_{29} -(D-5) d_{45} \cr
F_{;m} K^{ab} R_{ammb} & \hspace{1cm} 0= -(D-2) g_1 +4(D-1) g_6 +4(D-1) g_{13}
          +2g_{15} -(D-2) g_{20}\cr
   & \hspace{1cm} \phantom{0=} +2g_{22} -d_3
         -\frac 1 2 (D-2) d_{22}
          +(D-1) d_{26} +2d_{28}\cr
  &\hspace{1cm}  \phantom{0=}  +2d_{29} -(D-5) d_{46}\cr
F_{;m} K_{ab} K^{bc} K_c^a & \hspace{1cm} 0= (D-2) g_1 -4(D-1) g_6 -4(D-1) 
g_{13}
          -2g_{15} -d_3 -(D-2) d_{27}\cr
 & \hspace{1cm} \phantom{0=}  +d_{28}
            +2d_{29}
          -\frac 1 2 (D-2) d_{35} +(D-1) d_{39} +4d_{40} -(D-5) d_{51}\cr
FR_{ac} K^c_b K^{ab} & \hspace{1cm} 0= -2(D-2) g_8 -8g_9 +4g_{15} +g_{20} +2d_3
           +4d_{13} +4d_{14} -4d_{19}\cr
 & \hspace{1cm} \phantom{0=}  -(D-2) d_{27} +d_{28} +2d_{29}
\cr}
\eql{app2}
$$
Finally the third group:
$$\eqalign{
\underline{\mbox{Term}} & \hspace{1cm} \underline{\mbox{Coefficient}}\cr
FK_{:b}K^{:b} & \hspace{1cm} 0=2(D-1) g_6 -4g_{15} -(D-2) g_{20} +2g_{22}
             - \frac 1 2 (D-2) d_1 \cr
   &\hspace{1cm}  \phantom{0=} +2(D-1) d_2 +2d_{10}
              +d_{11}
               -(D-3) d_{15} -d_{16} -d_{18} +(D-2) g_{40}\cr
  &\hspace{1cm}  \phantom{0=}
              -4(D-1) g_{42} -2d_{23} -2d_{25}\cr
FK_{ab:}^{\phantom{ab:}a} K_:^b &\hspace{1cm}  0=2(D-2) g_1 -4(D-1) g_6 
-8(D-1) g_{13}
                +(4D-6) g_{20} -8g_{22}\cr
 & \hspace{1cm} \phantom{0=}  -(D-2) d_1
        +4(m-1) d_2 -(D-3) d_{11}
                +2d_{12} -2d_{14} +2d_{16}\cr
  & \hspace{1cm} \phantom{0=} -2d_{17} +2d_{18} -2(D-2) d_{25}
                +2d_{26} -4d_{29} \cr
FK_{ab:c} K^{ab\phantom{:}c}_{\phantom{ab}:} & \hspace{1cm} 0=(D-2) g_1 
-4(D-1) g_6
              - 4(D-1) g_{13} -2g_{15} +(D-2) g_{20}\cr
  &\hspace{1cm}  \phantom{0=}  -2g_{22} -3d_3
              +2d_{13} +2d_{14} -(D-3) d_{19} +(D-2) g_{41} \cr
&\hspace{1cm}  \phantom{0=}
               -4(D-1) g_{43} -2d_{24} -2d_{27}  \cr
FKK_{ab:}^{\phantom{ab:}ab} & \hspace{1cm} 0= 4(D-2) g_8 +16g_9 -4g _{15} 
-Dg_{20} +2g_{22}
                +d_{11} +2d_{12} -(D-4) d_{16}\cr
          &\hspace{1cm}  \phantom{0=} -2d_{17} -d_{18} -(D-2) d_{25}
                +d_{26} -2d_{29}\cr}
$$
$$\eqalign{
F_{;m} K_{:a}^{\phantom{:a} a} &\hspace{1cm}  0= -\frac 3 2 (D-2) g_1 
+4(D-1) g_6 -4(D-2) g_8
                -16g_9 +6(D-1) g_{13}\cr
  & \hspace{1cm} \phantom{0=}  +\frac 1 2 (D-2) d_1
          -2(D-1) d_2 -d_3
                -\frac 1 2 (D-2) d_4 +\frac 1 2 (D-2) d_6\cr
   & \hspace{1cm} \phantom{0=}-\frac 1 2 (D-2)
                 d_8 -2(D-1) d_{10}\cr
       & \hspace{1cm} \phantom{0=}  -d_{11} -2d_{13} +2(D-1) d_{15}
                 +d_{16} +d_{18} +2d_{19} -(D-5) d_{43}\cr
F_{;m} K_{ab:}^{\phantom{ab:}ab} &\hspace{1cm}  0= \frac 1 2 (D-2) g_1 
-2(D-1) g_6
             +4(D-2) g_8 +16g_9 -2(D-1) g_{13}\cr
   & \hspace{1cm} \phantom{0=}  +2g_{15} +2d_3
       -\frac 1 2
            (D-2) d_5 +\frac 1 2 (D-2) d_7 -\frac 1 2 (D-2) d_9 \cr
  & \hspace{1cm} \phantom{0=}
            -(D-1) d_{11} -2d_{12} -2d_{14}
          +(D-1) d_{16} +2d_{17} +
               (D-1) d_{18}\cr
    &\hspace{1cm}  \phantom{0=}-(D-5) d_{44}\cr
FK_{ab:}^{\phantom{ab:} a} K^{bc}_{\phantom{bc}:c} & 
\hspace{1cm} 0= (4-3D) g_{20}
                +6g_{22} -2d_3 -2(D-2) d_{12} -4d_{13} -2d_{14}\cr
    & \hspace{1cm} \phantom{0=}
          +(D+1) d_{17} +4d_{19} -(D-2) d_{27} +d_{28} +2d_{29}\cr
F K_{:ab} K^{ab} & \hspace{1cm} 0= 2(D-2) g_1 -4(D-1) g_6 -2(D-2) g_8 -8g_9
             -8 (D-1) g_{13} \cr
   & \hspace{1cm} \phantom{0=}+(4D-5) g_{20}
           -8g_{22} -(D-2) d_1
             +4(D-1) d_2 -(D-2) d_{11}\cr
   & \hspace{1cm} \phantom{0=} -2d_{14} +(D-2) d_{16}
            +3d_{18}
       -(D-2) d_{25} +d_{26} -2d_{29}
\cr}
\eql{app3}
$$
This completes the list of relations used for the calculation
of the $A_{5/2}$ coefficient.
\section{\bf References}
\vskip 5truept
\begin{putreferences}
\ref{APS}{Atiyah,M.F., V.K.Patodi and I.M.Singer,
\mpcps{77}{75}{43}.}
\ref{AandT}{Awada,M.A. and D.J.Toms: Induced gravitational and gauge-field 
actions from quantised matter fields in non-abelian Kaluza-Klein thory 
\np{245}{84}{161}.}
\ref{BandI}{Baacke,J. and Y.Igarishi: Casimir energy of confined massive 
quarks \prD{27}{83}{460}.}
\ref{Barnesa}{Barnes,E.W.: On the Theory of the multiple Gamma function 
{\it Trans. Camb. Phil. Soc.} {\bf 19} (1903) 374.}
\ref{Barnesb}{Barnes,E.W.: On the asymptotic expansion of integral 
functions of multiple linear sequence, {\it Trans. Camb. Phil. 
Soc.} {\bf 19} (1903) 426.}
\ref{Barv}{Barvinsky,A.O. Yu.A.Kamenshchik and I.P.Karmazin: One-loop 
quantum cosmology \aop {219}{92}{201}.}
\ref{BandM}{Beers,B.L. and Millman, R.S. :The spectra of the 
Laplace-Beltrami
operator on compact, semisimple Lie groups. \ajm{99}{1975}{801-807}.}
\ref{BandH}{Bender,C.M. and P.Hays: Zero point energy of fields in a 
confined volume \prD{14}{76}{2622}.}
\ref{BBG}{Bla\v zi\' c,N., Bokan,N. and Gilkey,P.B.: Spectral geometry of the 
form valued Laplacian for manifolds with boundary \ijpam{23}{92}{103-120}}
\ref{BEK}{Bordag,M., Elizalde,E. and Kirsten,K., 
\jmp{37}{96}{895}.}
\ref{BGKE}{Bordag,M., B.Geyer, K.Kirsten and E.Elizalde,: { Zeta function
determinant of the Laplace operator on the D-dimensional ball}, 
\cmp{179}{96}{215}.}
\ref{BKD}{Bordag,M., Kirsten,K. and Dowker,J.S.,
\cmp{182}{96}{371}.}
\ref{Branson}{Branson,T.P.: Conformally covariant equations on differential
forms \cpde{7}{82}{393-431}.}
\ref{BandG2}{Branson,T.P. and Gilkey,P.B. {\it Comm. Partial Diff. Eqns.}
{\bf 15} (1990) 245.}
\ref{BGV}{Branson,T.P., P.B.Gilkey and D.V.Vassilevich, 
Bollettino U. M. I. (7) 
{\bf 11}-B Suppl. Fasc. 2 (1997) 39.}
\ref{BCZ1}{Bytsenko,A.A, Cognola,G. and Zerbini, S. : Quantum fields in
hyperbolic space-times with finite spatial volume, hep-th/9605209.}
\ref{BCZ2}{Bytsenko,A.A, Cognola,G. and Zerbini, S. : Determinant of 
Laplacian on a non-compact 3-dimensional hyperbolic manifold with finite
volume, hep-th /9608089.}
\ref{CandH2}{Camporesi,R. and Higuchi, A.: Plancherel measure for $p$-forms
in real hyperbolic space, \jgp{15}{94}{57-94}.} 
\ref{CandH}{Camporesi,R. and A.Higuchi {\it On the eigenfunctions of the 
Dirac operator on spheres and real hyperbolic spaces}, gr-qc/9505009.}
\ref{ChandD}{Chang, Peter and J.S.Dowker :Vacuum energy on orbifold factors
of spheres, \np{395}{93}{407}.}
\ref{cheeg1}{Cheeger, J.,
\jdg {18}{83}{575}.}
\ref{cheeg2}{Cheeger,J.: Hodge theory of complex cones {\it Ast\'erisque} 
{\bf 101-102}(1983) 118-134}
\ref{Chou}{Chou,A.W.: The Dirac operator on spaces with conical 
singularities and positive scalar curvature, \tams{289}{85}{1-40}.}
\ref{CandT}{Copeland,E. and Toms,D.J.: Quantized antisymmetric 
tensor fields and self-consistent dimensional reduction 
in higher-dimensional spacetimes, \break\np{255}{85}{201}}
\ref{DandH}{D'Eath,P.D. and J.J.Halliwell: Fermions in quantum cosmology 
\prD{35}{87}{1100}.}
\ref{cheeg3}{Cheeger,J.:Analytic torsion and the heat equation. \aom{109}
{79}{259-322}.}
\ref{DandE}{D'Eath,P.D. and G.V.M.Esposito: Local boundary conditions for 
Dirac operator and one-loop quantum cosmology \prD{43}{91}{3234}.}
\ref{DandE2}{D'Eath,P.D. and G.V.M.Esposito: Spectral boundary conditions 
in one-loop quantum cosmology \prD{44}{91}{1713}.}
\ref{Dow1}{Dowker,J.S.: Effective action on spherical domains, \cmp{162}{94}
{633}.}
\ref{Dow8}{Dowker,J.S. {\it Robin conditions on the Euclidean ball} 
MUTP/95/7; hep-th\break/9506042. {\it Class. Quant.Grav.} to be published.}
\ref{Dow9}{Dowker,J.S. {\it Oddball determinants} MUTP/95/12; 
hep-th/9507096.}
\ref{Dow10}{Dowker,J.S. {\it Spin on the 4-ball}, 
hep-th/9508082, {\it Phys. Lett. B}, to be published.}
\ref{DandA2}{Dowker,J.S. and J.S.Apps, {\it Functional determinants on 
certain domains}. To appear in the Proceedings of the 6th Moscow Quantum 
Gravity Seminar held in Moscow, June 1995; hep-th/9506204.}
\ref{DABK}{Dowker,J.S., Apps,J.S., Bordag,M. and Kirsten,K.: Spectral 
invariants for the Dirac equation with various boundary conditions 
{\it Class. Quant.Grav.} to be published, hep-th/9511060.}
\ref{EandR}{E.Elizalde and A.Romeo : An integral involving the
generalized zeta function, {\it International J. of Math. and 
Phys.} {\bf13} (1994) 453.}
\ref{ELV2}{Elizalde, E., Lygren, M. and Vassilevich, D.V. : Zeta function 
for the laplace operator acting on forms in a ball with gauge boundary 
conditions. hep-th/9605026}
\ref{ELV1}{Elizalde, E., Lygren, M. and Vassilevich, D.V. : Antisymmetric
tensor fields on spheres: functional determinants and non-local
counterterms, \jmp{}{96}{} to appear. hep-th/ 9602113}
\ref{Kam2}{Esposito,G., A.Y.Kamenshchik, I.V.Mishakov and G.Pollifrone: 
Gravitons in one-loop quantum cosmology \prD{50}{94}{6329}; 
\prD{52}{95}{3457}.}
\ref{Erdelyi}{A.Erdelyi,W.Magnus,F.Oberhettinger and F.G.Tricomi {\it
Higher Transcendental Functions} Vol.I McGraw-Hill, New York, 1953.}
\ref{Esposito}{Esposito,G.: { Quantum Gravity, Quantum Cosmology and 
Lorentzian Geometries}, Lecture Notes in Physics, Monographs, Vol. m12, 
Springer-Verlag, Berlin 1994.}
\ref{Esposito2}{Esposito,G. {\it Nonlocal properties in Euclidean Quantum
Gravity}. To appear in Proceedings of 3rd. Workshop on Quantum Field Theory
under the Influence of External Conditions, Leipzig, September 1995; 
gr-qc/9508056.}
\ref{EKMP}{Esposito G, Kamenshchik Yu A, Mishakov I V and Pollifrone G.:
One-loop Amplitudes in Euclidean quantum gravity.
\prd {52}{96}{3457}.}
\ref{ETP}{Esposito,G., H.A.Morales-T\'ecotl and L.O.Pimentel {\it Essential
self-adjointness in one-loop quantum cosmology}, gr-qc/9510020.}
\ref{FORW}{Forgacs,P., L.O'Raifeartaigh and A.Wipf: Scattering theory, U(1) 
anomaly and index theorems for compact and non-compact manifolds 
\np{293}{87}{559}.}
\ref{GandM}{Gallot S. and Meyer,D. : Op\'erateur de coubure et Laplacian
des formes diff\'eren-\break tielles d'une vari\'et\'e riemannienne 
\jmpa{54}{1975}
{289}.}
\ref{Gilkey1}{Gilkey, P.B, Invariance theory, the heat equation and the
Atiyah-Singer index theorem, 2nd. Edn., CRC Press, Boca Raton 1995.}
\ref{Gilkey2}{Gilkey,P.B.:On the index of geometric operators for 
Riemannian manifolds with boundary \aim{102}{93}{129}.}
\ref{Gilkey3}{Gilkey,P.B.: The boundary integrand in the formula for the 
signature and Euler characteristic of a manifold with boundary 
\aim{15}{75}{334}.}
\ref{Grubb}{Grubb,G. {\it Comm. Partial Diff. Eqns.} {\bf 17} (1992) 
2031.}
\ref{GandS1}{Grubb,G. and R.T.Seeley \cras{317}{1993}{1124}; \invm{121}{95}
{481}.}
\ref{GandS}{G\"unther,P. and Schimming,R.:Curvature and spectrum of compact
Riemannian manifolds, \jdg{12}{77}{599-618}.}
\ref{IandT}{Ikeda,A. and Taniguchi,Y.:Spectra and eigenforms of the 
Laplacian
on $S^n$ and $P^n(C)$. \ojm{15}{1978}{515-546}.}
\ref{IandK}{Iwasaki,I. and Katase,K. :On the spectra of Laplace operator
on $\La^*(S^n)$ \pja{55}{79}{141}.}
\ref{JandK}{Jaroszewicz,T. and P.S.Kurzepa: Polyakov spin factors and 
Laplacians on homogeneous spaces \aop{213}{92}{135}.}
\ref{Kam}{Kamenshchik,Yu.A. and I.V.Mishakov: Fermions in one-loop quantum 
cosmology \prD{47}{93}{1380}.}
\ref{KandM}{Kamenshchik,Yu.A. and I.V.Mishakov: Zeta function technique for
quantum cosmology {\it Int. J. Mod. Phys.} {\bf A7} (1992) 3265.}
\ref{KandC}{Kirsten,K. and Cognola.G,: { Heat-kernel coefficients and 
functional determinants for higher spin fields on the ball} \cqg{13}{96}
{633-644}.}
\ref{Levitin}{Levitin,M.: { Dirichlet and Neumann invariants for Euclidean
balls}, {\it Diff. Geom. and its Appl.}, to be published.}
\ref{Luck}{Luckock,H.C.: Mixed boundary conditions in quantum field theory 
\jmp{32}{91}{1755}.}
\ref{MandL}{Luckock,H.C. and Moss,I.G,: The quantum geometry of random 
surfaces and spinning strings \cqg{6}{89}{1993}.}
\ref{Ma}{Ma,Z.Q.: Axial anomaly and index theorem for a two-dimensional 
disc 
with boundary \jpa{19}{86}{L317}.}
\ref{Mcav}{McAvity,D.M.: Heat-kernel asymptotics for mixed boundary 
conditions \cqg{9}{92}{1983}.}
\ref{MandV}{Marachevsky,V.N. and D.V.Vassilevich {\it Diffeomorphism
invariant eigenvalue \break problem for metric perturbations in a bounded 
region}, SPbU-IP-95, \break gr-qc/9509051.}
\ref{Milton}{Milton,K.A.: Zero point energy of confined fermions 
\prD{22}{80}{1444}.}
\ref{MandS}{Mishchenko,A.V. and Yu.A.Sitenko: Spectral boundary conditions 
and index theorem for two-dimensional manifolds with boundary 
\aop{218}{92}{199}.}
\ref{Moss}{Moss,I.G.,
\cqg{6}{89}{759}.}
\ref{MandP}{Moss,I.G. and S.J.Poletti: Conformal anomaly on an Einstein space 
with boundary \pl{B333}{94}{326}.}
\ref{MandP2}{Moss,I.G. and S.J.Poletti \np{341}{90}{155}.}
\ref{NandOC}{Nash, C. and O'Connor,D.J.: Determinants of Laplacians, the 
Ray-Singer torsion on lens spaces and the Riemann zeta function 
\jmp{36}{95}{1462}.}
\ref{NandS}{Niemi,A.J. and G.W.Semenoff: Index theorem on open infinite 
manifolds \np {269}{86}{131}.}
\ref{NandT}{Ninomiya,M. and C.I.Tan: Axial anomaly and index thorem for 
manifolds with boundary \np{245}{85}{199}.}
\ref{norlund2}{N\"orlund~N. E.:M\'emoire sur les polynomes de Bernoulli.
\am {4}{21} {1922}.}
\ref{Poletti}{Poletti,S.J. \pl{B249}{90}{355}.}
\ref{RandT}{Russell,I.H. and Toms D.J.: Vacuum energy for massive forms 
in $R^m\times S^N$, \cqg{4}{86}{1357}.}
\ref{RandS}{R\"omer,H. and P.B.Schroer \pl{21}{77}{182}.}
\ref{Trautman}{Trautman,A.: Spinors and Dirac operators on hypersurfaces 
\jmp{33}{92}{4011}.}
\ref{Vass}{Vassilevich,D.V.{Vector fields on a disk with mixed 
boundary conditions} gr-qc /9404052.}
\ref{Voros}{Voros,A.:
Spectral functions, special functions and the Selberg zeta function.
\cmp{110}{87}439.}
\ref{Ray}{Ray,D.B.: Reidemeister torsion and the Laplacian on lens
spaces \aim{4}{70}{109}.}
\ref{McandO}{McAvity,D.M. and Osborn,H., 
\cqg{8}{91}{1445}.}
\ref{AandE}{Avramidi,I. and Esposito,G., 
Class. Quauntum Grav. {\bf 15} (1998) 281}
\ref{AandE1}{Avramidi,I. and Esposito,G., Lack of strong
ellipticity in Euclidean quantum gravity, hep-th/9708163.}
\ref{AandE2}{Avramidi,I. and Esposito,G., Gauge theories on
manifolds with boundary, hep-th/9710048.} 
\ref{MandS}{Moss,I.G. and Silva P.J., Invariant boundary conditions for
gauge theories gr-qc/9610023.}
\ref{barv}{Barvinsky,A.O.\pl{195B}{87}{344}.}
\ref{krantz}{Krantz,S.G. Partial Differential Equations and Complex
Analysis (CRC Press, Boca Raton, 1992).}
\ref{treves}{Treves,F. Introduction to Pseudodifferential and Fourier Integral
Operators,\break Vol.1, (Plenum Press,New York,1980).}
\ref{EandS}{Egorov,Yu.V. and Shubin,M.A. Partial Differential Equations
(Springer-Verlag, Berlin,1991).}
\ref{AandS}{Abramowitz,M. and Stegun,I.A. Handbook of Mathematical Functions 
(Dover, New York, 1972).}
\ref{ACNY}{Abouelsaood,A., Callan,C.G., Nappi,C.R. and Yost,S.A.\np{280}{87}
{599}.}
\ref{BGKE}{Bordag,M., B.Geyer, K.Kirsten and E.Elizalde, { Zeta function
determinant of the Laplace operator on the D-dimensional ball}, 
\cmp{179}{96}{215}.}
\ref{ivan91}{Avramidi, I.G., \np{355}{91}{712}.}
\ref{fulken88}{Fulling, S.A. and Kennedy, G., 
{\it Trans. Amer. Math. Soc.} {\bf 310} (1988) 583.}
\ref{amster89}{Amsterdamski, P., Berkin, A.L. and O'Connor, D.J., \cqg{6}{89}
{1981}.}
\ref{osb91}{Mc Avity, D.M. and Osborn, H., \cqg{8}{91}{603}.}
\ref{and92}{Dettki, A. and Wipf, A., \np{377}{92}{252}.}
\ref{trento90}{Cognola, G., Vanzo, L. and Zerbini, S., \pl{241}{90}{381}.} 
\ref{olver54}{Olver, G.W., {\it Phil. Trans. Roy. Soc.} {\bf A247} (1954)
328.}
\ref{watson}{Watson,G.N., Theory of Bessel Functions (Cambridge University
Press, Cambridge,1944.}
\ref{GandR}{Gradshteyn, I.S. and Ryzhik, I.M., Tables of Integrals, Series
and Products (Academic Press, New York, 1965).}
\ref{HandE}{Hawking, S.W. and Ellis, G.F.R., The large scale structure of
space-time (Cambridge University Press, Cambridge, 1973).}
\ref{DK}{Dowker, J.S. and Kirsten, K., \cqg{14}{97}{L169}.}
\ref{BGG}{Branson, T., Gilkey, P. and Grubb, G., private communication.}
\ref{MandD}{Moss, I.G. and Dowker, J.S., \pl{229}{89}{261}.}
\ref{DandS}{Dowker, J.S. and Schofield, J.P., \jmp{31}{90}{808}.}
\ref{ven97}{Van de Ven, A., {\it Index-free Heat Kernel Coefficients}, 
hep-th/9708152.}
\ref{KK}{Kirsten, K., Class. Quantum Grav. {\bf 15} (1998) L5.}
\end{putreferences} 
\bye